\documentclass[10pt,journal,compsoc]{IEEEtran}
\ifCLASSOPTIONcompsoc
  \usepackage[nocompress]{cite}
\else
  \usepackage{cite}
\fi
\ifCLASSINFOpdf
\else
\fi
\usepackage{cite}
\usepackage{amsmath,amssymb,amsfonts}
\usepackage{graphicx}
\usepackage{listings}
\usepackage{textcomp}
\usepackage{xcolor}
\usepackage{subfigure}
\usepackage[noend]{algpseudocode}
\usepackage{algorithmicx,algorithm}
\usepackage{algcompatible}
\usepackage{threeparttable}
\usepackage{tikz}
\usetikzlibrary{fit, calc}
\usetikzlibrary{decorations}
\usepackage{verbatim}
\usepackage{pgfplotstable}
\usepackage{diagbox}
\usepackage{makebox}
\usepackage{multirow}
\usepackage{makecell}
\usepackage{booktabs}
\usepackage[numbers,sort&compress]{natbib} 
\usepackage[justification=centering]{caption}
\usepackage{float}
\usepackage{graphicx}

\usepackage{multirow}

\usepackage{subcaption}

\begin{document}
%
\title{An Adversarial Robust Behavior Sequence Anomaly Detection Approach Based on Critical Behavior Unit Learning}
%
%
%
%

\author{Dongyang Zhan,~\IEEEmembership{Member,~IEEE, }
        Kai Tan,
        Lin Ye*,
        Xiangzhan Yu, 
        Hongli Zhang, 
        and~Zheng He
\IEEEcompsocitemizethanks{\IEEEcompsocthanksitem D. Zhan, K. Tan, X. Yu, H. Zhang, L. Ye are with the School of Cyberspace Science, Harbin Institute of Technology, Harbin,
Heilongjiang, 150001.\\
Z. He is with the Heilongjiang Meteorological Bureau, Harbin, Heilongjiang, 150001.
\protect\\
E-mail: \{zhandy, yuxiangzhan, zhanghongli, hityelin\}@hit.edu.cn
\IEEEcompsocthanksitem * Corresponding Author: hityelin@hit.edu.cn}
}

\markboth{Journal of \LaTeX\ Class Files,~Vol.~14, No.~8, August~2015}%
{Shell \MakeLowercase{\textit{et al.}}: Bare Demo of IEEEtran.cls for Computer Society Journals}
%



\IEEEtitleabstractindextext{%
\begin{abstract}
Sequential deep learning models (e.g., RNN and LSTM) can learn the sequence features of software behaviors, such as API or syscall sequences. However, recent studies have shown that these deep learning-based approaches are vulnerable to adversarial samples. Attackers can use adversarial samples to change the sequential characteristics of behavior sequences and mislead malware classifiers. In this paper, an adversarial robustness anomaly detection method based on the analysis of behavior units is proposed to overcome this problem. We extract related behaviors that usually perform a behavior intention as a behavior unit, which contains the representative semantic information of local behaviors and can be used to improve the robustness of behavior analysis. By learning the overall semantics of each behavior unit and the contextual relationships among behavior units based on a multilevel deep learning model, our approach can mitigate perturbation attacks that target local and large-scale behaviors. In addition, our approach can be applied to both low-level and high-level behavior logs (e.g., API and syscall logs). The experimental results show that our approach outperforms all the compared methods, which indicates that our approach has better performance against obfuscation attacks.
\end{abstract}

}

\maketitle

\IEEEdisplaynontitleabstractindextext

%
\IEEEpeerreviewmaketitle

\IEEEraisesectionheading{\section{Introduction}\label{sec:introduction}}

%
%
%
%

\IEEEPARstart{T}{he} amount of malware is rapidly growing. 
Malware such as ransomware and Trojans are also rapidly changing and have become the most serious threats in cyberspace\cite{afianian2019malware}. Therefore, detecting malware is very important for cybersecurity.

To improve the efficiency of malware detection, security researchers have proposed many detection methods based on machine/deep learning technology. Deep neural networks have been shown to enable efficient and accurate malware classification\cite{sahin2020survey}. Existing machine/deep learning-based malware detection and classification systems mainly learn features by static analysis of executable files\cite{amin2020static} or dynamic behavior analysis\cite{li2022novel,or2019dynamic}. However, static analysis methods are insufficient code adversarial technologies (e.g., code obfuscation, dynamic code loading, and shelling)\cite{or2019dynamic}. In contrast, dynamic analysis approaches \cite{or2019dynamic} can overcome these problems by tracking and analyzing the execution processes of the target programs. Specifically, these approaches usually leverage virtual environments or kernel modules to trace program execution and record sequential behavior logs (e.g., APIs or syscalls). During the execution of a program/system, it will trigger many events (e.g., syscalls and APIs), which can be intercepted or recorded for security analysis. We define these events as behaviors. There are many dynamic analysis approaches proposed by security researchers, such as rule-based approaches \cite{forrest1996sense} and machine-learning-based approaches \cite{firdausi2010analysis}. 

In recent years, combining dynamic approaches and deep learning techniques has risen. These approaches can automatically learn features and train neural networks from behavioral sequences. Sequential neural networks (e.g., RNN-based models and transformers) are usually employed to perform sequence anomaly detection \cite{vinayakumar2018detecting} based on dynamic behavior logs and can be used for malware detection, since these deep learning approaches have a good ability to model sequential data.

However, these sequential neural network models are vulnerable to adversarial sample attacks\cite{liang2022adversarial}, which perturb the samples of the model's input to bypass the detection of these models. Unlike adversarial samples for images \cite{fang2022novel}, the adversarial samples for dynamic malware detection are behavior logs. The adversarial samples need to be practical, so attackers cannot directly replace/remove the original behaviors of the malware. Therefore, the generation approaches of adversarial samples mainly insert irrelevant behavior sequences (i.e., normal behavior fragments) into the original behavior sequences or replace the anomalous behaviors with those with similar functions \cite{hu2018black,fadadu2019evading,rosenberg2020query}. For instance, \cite{rosenberg2020query} can achieve up to 87.94\% for an LSTM classifier by injecting perturbed behaviors.

To improve the robustness of malware detection, many approaches are proposed to analyze and learn high-level OS-related characteristics of program behaviors. For instance, DroidSpan \cite{cai2020assessing} learns the behavior characteristics associated with accessing Android sensitive data (e.g., user accounts), which are found to change little with the evolution of Android software. Droidcat \cite{cai2018droidcat} captures lots of app-level dynamic characteristics of Android apps to perform robust malware identification. However, these approaches are usually based on the high-level semantics of APIs in specific operating systems, such as the APIs for accessing accounts in Android, which is the key to achieving robust analysis. So, it is not easy to apply these approaches to low-level behavior logs that do not have such high-level semantic information (e.g., Linux syscalls). 

To analyze behavior logs without high-level OS-related semantic information (e.g., syscall logs), many approaches for sequential data have been proposed, such as adversarial training \cite{szegedy2013intriguing}, defense Sequence-GAN \cite{rosenberg2019defense}, and sequence squeezing \cite{rosenberg2021sequence}. However, adversarial training reduces the detection performance of networks on unperturbed data samples. Defense Sequence-GAN suffers from the problem of training overhead and cannot handle adversarial samples with long sequence injections. Sequence squeezing can only defend against the attack of replacing labeled malicious behaviors with less well-known behaviors with similar functionality. In addition, these approaches lack in-depth analysis of the relationships among behaviors and therefore cannot solve the problems of adversarial attacks in the behavior analysis domain.

In this paper, an adversarial robust behavior sequence anomaly detection approach based on critical behavior unit learning is proposed. Based on our observation, the behavior of a program is composed of a series of representative behavior units. A behavior unit is a collection of related behaviors that are usually composed of multiple behaviors to perform a specific behavior purpose and contain the representative semantics of local behaviors. For instance, open, read, and close can constitute a behavior unit, and the purpose of this unit is to access a file. And, behavior units such as accessing files, sending data to the network, etc. can constitute the behavior of a program. 
Based on this observation, our approach learns the overall semantics of each behavior unit and the contextual relationships among behavior units to perform adversarial robust anomaly detection. By extracting and analyzing the behavior units, the overall representation and behavior intention of local behaviors can be obtained, and the robustness of behavior analysis can be improved. It is difficult to change the local behavior intention with small-scale behavior injections or replacements, so our approach is resistant to perturbation attacks that target local behaviors. Then, the contextual relationships among behavior units are analyzed to obtain the global representation of the target behaviors and mitigate perturbation attacks that target a wide range of behaviors. By combining the local and global features of target behaviors based on behavior unit analysis, our approach can perform robust anomaly detection for both high-level behavior logs (e.g., Android APIs) and low-level behavior logs (e.g., syscalls). 

Our approach first identifies behavior units from unperturbed behavior sequences, which contain the representative semantics of the original behavior sequences. Then, we extract such behavior units from perturbed samples for behavior analysis, since our goal is to analyze the security of behavior sequences with adversarial attacks. Last, we design a multilevel deep learning model to perform security analysis based on behavior units. Even if an attacker tampers with some critical behaviors or behavior units, which the attacker needs to replace with behaviors of the same function to keep the program's functionality, this threat can be identified by our approach.

In summary, the first contribution of this paper is that an adversarial robust behavior sequence anomaly detection approach is proposed based on the analysis of behavior units. 
By learning the overall semantics of each behavior unit and the contextual relationships among behavior units, our model can improve the robustness of behavior analysis and can be applied to both low-level and high-level behavior logs.


The second contribution is that after implementing the prototype, comparative experiments are carried out to indicate the performance and robustness of our approach and the importance of the behavior unit feature.

The rest of this paper is organized as follows. Section \ref{s:related-work} summarizes the related work. The threat model and defense strategy are introduced in Section \ref{s:Threat}. Section \ref{s:design} describes the overall framework and its implementation. The evaluation of our approach is performed in Section \ref{s:evaluation}. Section \ref{s:conclusion} summarizes this paper. 

\section{Related Work}\label{s:related-work}
In this section, we summarize related work about deep learning-based malware detection, adversarial attacks of sequential models and adversarial defense approaches.

\subsection{Deep Learning-based Malware Detection}

Detecting malware is a hot topic in computer security. Compared with static analysis approaches\cite{amin2020static}, dynamic analysis approaches\cite{li2022novel,hu2021using} are robust for adversarial technologies (e.g., code obfuscation, dynamic code loading, and shelling)\cite{or2019dynamic}. With the development of deep learning, leveraging deep learning to detect malware by analyzing the behaviors of malware is an important approach in malware detection\cite{hasan2021megdroid}.  

Deep learning-based models can analyze dynamic execution information (e.g., APIs and syscalls) of software to identify anomalies. The invoked API sequences can be effectively applied to model the most representative behavior features of malware \cite{d2021association}. For example, \cite{hardy2016dl4md,rhode2018early,natani2013malware} use sequences of API calls for malware detection. These approaches are based on frequency analysis of API calls \cite{natani2013malware}, or identify specific malicious API call sequence characteristics \cite{hardy2016dl4md,rhode2018early}. 
Therefore, sequences of API calls can be effectively employed to model the most representative behavioral features associated with particular malware applications. 

Since sequential deep learning models have a strong ability to learn sequential features, many approaches use sequential deep learning models to detect abnormal behaviors of malware. \cite{xiao2019android} proposes an LSTM-based detection approach. This approach converts system call events to semantic information in natural language, and treats a system call event as a sentence. Then, an LSTM-based classifier is proposed to identify anomalies. \cite{guan2021malware} proposes an LSTM-transformer architecture to improve the classification of malicious system calls, which leverages the ability of LSTM to capture sequential pattern features and the ability of the transformer-encoder to capture global dependencies. This approach combines the strengths of these two models to identify abnormal patterns in system calls.

\subsection{Adversarial Attacks of Sequential Models}

Sequential neural network models are vulnerable to adversarial sample attacks. Different from adversarial sample attacks in the image recognition field, attackers can inject irrelevant behavior sequences into the original behavior sequences or modify some unimportant behaviors to generate adversarial samples. Deleting or modifying the behaviors of malware may affect the functionality of the malware. Therefore, these approaches usually generate perturbed samples by inserting intercepted benign fragments or generating fragments, which significantly reduces the ability of sequential neural network-based methods\cite{hu2018black,fadadu2019evading,rosenberg2020query}. For instance, \cite{rosenberg2018generic} proposes an end-to-end black-box method to generate adversarial examples for RNN-based models by changing unnecessary features. \cite{fadadu2019evading} mimics benign behaviors in the malware by periodically injecting intelligently selected API calls in original malicious API call sequences, to bypass the classifiers. 
\cite{hu2018black} outputs sequential adversarial examples based on a generative RNN and injects them into malicious sequences to attack RNN-based malware detection systems. \cite{rosenberg2020query} perturbs the classifier by injecting normal sequences into abnormal sequences. Since directly injecting benign sequences into malicious sequences is easy to identify, Sequence-GAN is used to generate benign sequences. An algorithm is proposed to minimize the amount of injection. Furthermore, by injecting enough benign sequences, perturbed samples can completely bypass the classifier.

\subsection{Adversarial Defense Approaches}

To defend against such adversarial attacks, several adversarial defense approaches have been proposed. Adversarial learning\cite{szegedy2013intriguing} is a common approach to defend against adversarial sample attacks, which improves model robustness by adding adversarial examples to the training set. However, adversarial training has several limitations. First, the robustness of the model heavily depends on the quality of adversarial samples. Second, if the generated perturbed samples are very similar to normal samples, it may reduce the model detection ability. Furthermore, the generality of new adversarial attacks is limited \cite{madry2017towards}.

Another defense approach is Defense Sequence-GAN\cite{rosenberg2019defense}, which filters out the perturbations added by adversarial attacks by training a GAN to model the distribution of unperturbed input. However, there are several problems with this approach. First, the training overhead of this approach is high, as discussed in \cite{rosenberg2019defense}. Second, this approach can only slightly mitigate adversarial attacks. This approach is not good for adversarial sample detection in long sequences, that is, if an attacker inserts a long normal sequence among a few malicious behavior sequences, it is difficult for this approach to detect anomalies.

Sequential squeezing\cite{rosenberg2021sequence} is a possible defense approach that merges similar behaviors into a single representative feature. This approach mainly defends against critical behavior obfuscation attacks, which replace well-known malicious behaviors with less well-known behaviors with similar functionality. However, this approach cannot be applied to identifying perturbations with benign behaviors.

Therefore, existing adversarial defense approaches cannot effectively defend against such adversarial attacks in the malicious behavior detection domain.

\section{Threat Model and Defense Strategies}\label{s:Threat}
In this section, we introduce the threat model and defense strategies of our approach.

\subsection{Threat Model}
Recent studies have shown that deep learning-based malware detection methods are vulnerable to adversarial attacks. The adversarial attack modifies the behavior execution sequence of malware so that the modified sequence is incorrectly classified as benign. 

We assume that attackers can only obtain limited knowledge of the target classification model by query-only observation, because it is usually difficult to obtain perfect details of the model architecture that are highly protected \cite{apruzzese2022real}. As discussed in \cite{rosenberg2020query,rosenberg2021sequence}, an attacker cannot modify behavior logs directly because they are generated by tracking program execution, which is different from directly modifying pixel data to perturb the image samples. To maintain the program’s functionality, attackers can only replace the original behaviors with functionally similar behaviors or insert irrelevant behaviors into the original sequences.

\subsection{Our Defense Strategies}

To defend against the above attacks, the defense strategies are as follows:

1) Strongly correlated behavior subsequences with obvious behavioral intentions are identified, and then the unrecognized irrelevant behaviors are excluded during the sequence detection process. According to our observation of syscall behaviors, we find that a collection of strongly related behaviors, which usually have a specific behavioral intention, constitute a behavioral unit. For instance, the operation of reading a file contains the syscall set of ``open, read, close". These behaviors are possible related syscalls, which have an obvious behavioral intention. If we can automatically identify these key behavioral intentions and delete irrelevant behaviors, we can defend against the interference of adversarial attacks.

2) The joint feature representations within and between behavior units are learned to identify the latent features of malicious behavior intentions and the dependence among multiple behavior intentions. The intention of some behavior units can clearly distinguish whether they are benign or malicious. However, we argue that multiple behavioral intentions can also further describe anomalies. 
To correlate the behavior units and learn the joint feature representations, a multilevel deep learning model based on the transformer encoder is proposed, as detailed in Section 4. First, a transformer encoder block is used to learn the embedded representation of each behavior unit. Next, we concatenate the embeddings of behavior units with the embeddings of the in-unit corresponding behaviors to generate the joint embeddings of behavior units and behaviors, which are fed into other transformer encoder blocks to learn the contextual and joint feature representations of the behavior and behavior unit sequences.

\section{System Design}\label{s:design}

This section describes the overall architecture of the proposed malicious behavior detection approach based on critical behavior unit learning. 

\subsection{Framework}

\begin{figure*}
\centering
\includegraphics[scale=0.6]{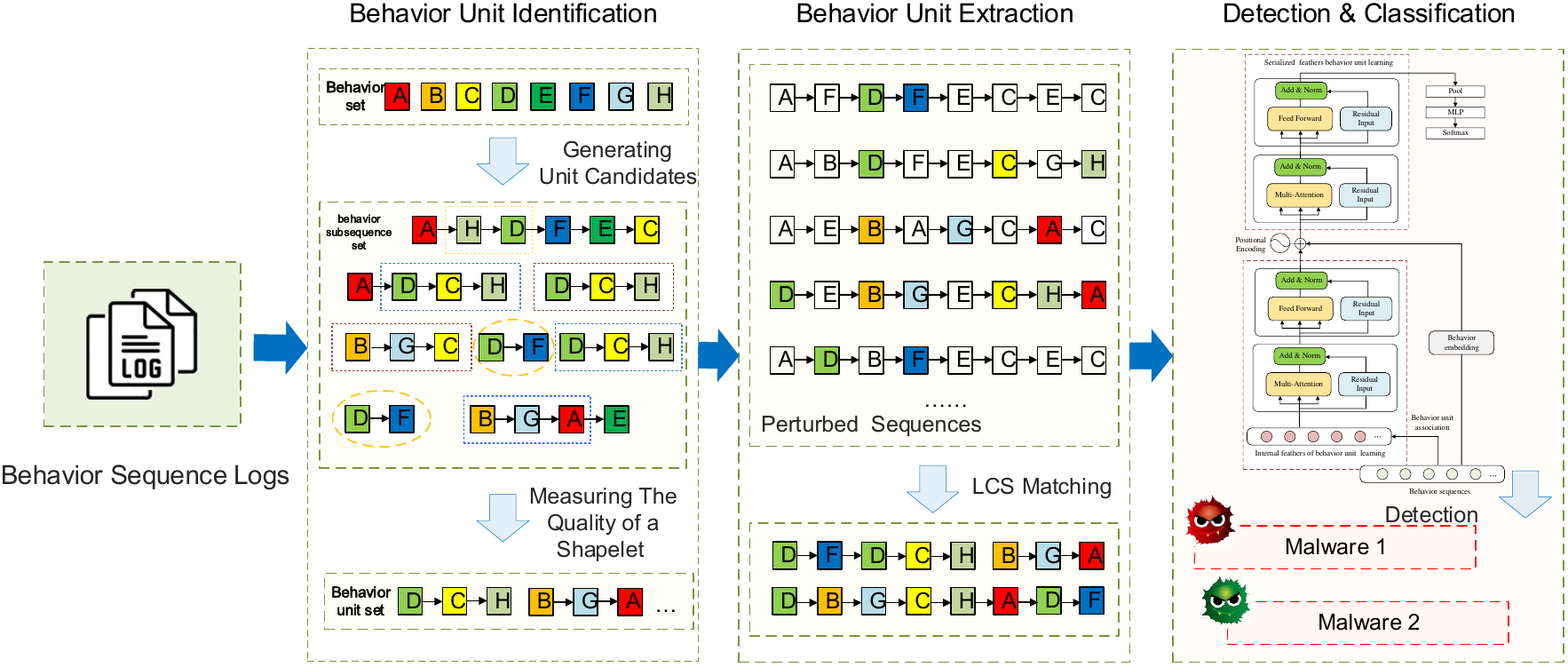}
\caption{\centering{The framework of our approach.}}
\label{fig_frame}
\end{figure*}

As shown in Figure \ref{fig_frame}, the input of our proposed system is behavior logs, which consists of three modules: (1) behavior unit pattern identification, (2) behavior unit extraction, and (3) feature extraction and behavior classification.

\textbf{1) Behavior Unit Pattern Identification}

This module collects the behavior sequences of normal and malicious software. Then, we identify the behavior subsequences with obvious behavior classification characteristics from the unperturbed behavior sequences. Based on the patterns of the identified behavior subsequences, we identify the behavior units in the perturbed sequences. Specifically, we obtain candidate subsequences of possible behaviors from unperturbed sequences, and then apply the shapelet algorithm to select subsequences with obvious classification characteristics as the pattern of the critical behavior unit, as described in Section \ref{s:pattern}.

\textbf{2) Behavior Unit Extraction}

To extract behavior units from the behavior sequences with perturbation and remove behaviors unrelated to critical behaviors, the module applies the long common subsequence (LCS) algorithm to extract behavior units from the perturbed behavior sequence based on the extracted patterns, which improves the robustness against obfuscation attacks. 

\textbf{3) Feature Extraction and Behavior Classification}

\begin{figure}
\centering
\includegraphics[scale=0.5]{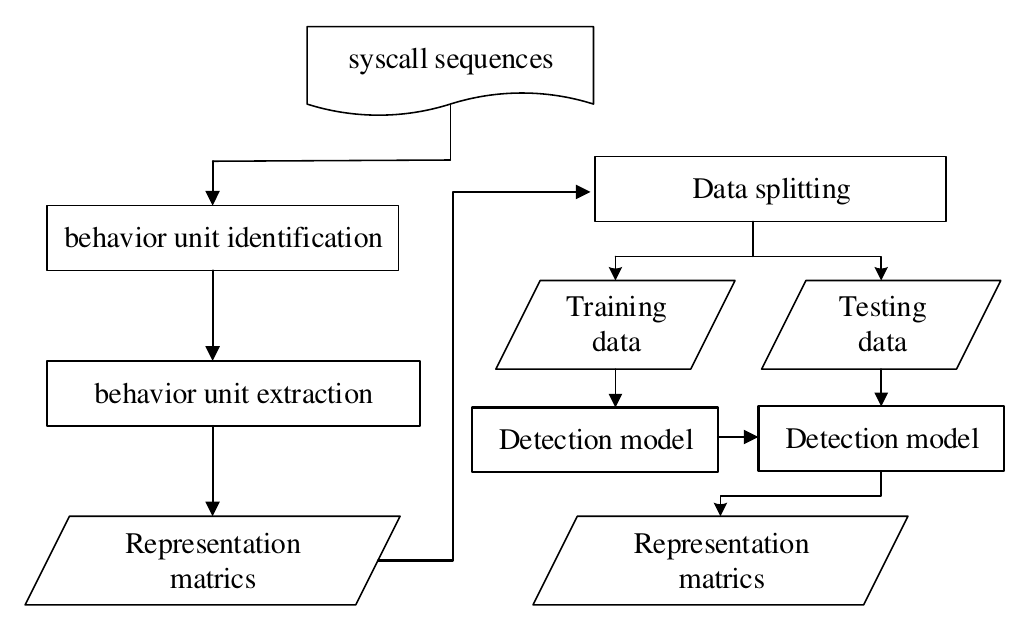}
\caption{\centering{The workflow of the proposed approach.}}
\label{fig_flow}
\end{figure}

This module integrates multilayer transformer models to extract multilayer features of software behaviors. The classification result determines whether the input sequence is normal or abnormal. Its workflow is shown in Figure \ref{fig_flow}. First, the behavior sequence is subjected to behavioral unit extraction, so perturbations from sequences of unrelated behaviors are excluded. Second, representations of the input behavior sequences, which are split into training and testing data, are generated. Last, the training data are used to train the transformer-based classification model, and the test data are fed into the trained transformer-based model to test the model performance.

\subsection{Behavior Unit Pattern Identification}\label{s:pattern}
As illustrated in Figure \ref{fig_frame}, this module generates behavioral sequence fragments from the unperturbed behavior sequences as behavior unit candidates and then evaluates the quality of these fragments to choose the top K fragments as critical behavior units.

\subsubsection{Behavior Unit Candidate Generation}\label{s:candidate}
We use binary classification to identify normal and abnormal behaviors with class labels $Y=\left\{{0,1}\right\}$ for the given unperturbed behavior sequences $I=\{ {I_1},{I_2}, \ldots {I_n}\}$. We aim to identify the critical behavior sequence fragments with obvious behavior features and extract the most representative fragment of all the behavior sequences to detect abnormal sequences.

We assume that $S$ is a subsequence/fragment in a behavior sequence ${I_i}$ and that the lengths are $l$ and $m$, respectively, where $l \le m$. Any behavior sequence of length $m$ contains $m - l + 1$ distinct subsequences of length $l$.  We denote the set of all subsequences of length $l$ for sequence ${I_i}$ to be ${W_{i,l}}$  and denote the set of all subsequences of length $l$ for the dataset to be
\begin{equation}
{W_l} = \{ {W_{1,l}},{W_{2,l}}, \ldots {W_{n,l}}\} 
\end{equation}

The set of all candidate critical behavior units for dataset $I$ is 
\begin{equation}
W = \{ {W_{min}},{W_{min + 1}}, \ldots {W_{max}}\}
\end{equation}
where $\min  \ge 3$  and $\max  \le m$ , and the process of extracting the top $k$ critical behavior units is defined in Algorithm 1.

\begin{algorithm}[h] 
\caption{Behavior Unit Candidate Extraction} 
\label{alg1} 
\begin{algorithmic}[1] 
\REQUIRE $I$ , $min$, $max$, $k$
\STATE${k\_behavior\_units=\emptyset}$
\STATE $C = {\rm{ }}classLabels(I);$
\FOR{ $behavior sequences$ ${I_i}$ in $I$}
\STATE${behavior\_units=\emptyset}$
\FOR{$l=min$ to $max$}
\STATE ${W_{i,l}} = generateCandidates({I_i},{\rm{ }}min,{\rm{ }}max);$
\FOR{ all $subsequence$ $S$ in ${W_{i,l}}$}
\STATE $quality{\rm{ }} = {\rm{ }}assessCandidate\left( S \right);$
\STATE $behavior\_units.add{\rm{ }}\left( {S,{\rm{ }}quality} \right);$
\ENDFOR
\ENDFOR
\STATE $sortByQuality\left( {behavior\_units} \right);$
\STATE $removeSelfSimilar\left( {behavior\_units} \right);$
\STATE $k\_behavior\_units{\rm{ }} = $
\Statex \;\;\,
$ {\rm{ }}$merge$\left( {k,{\rm{ }}k\_behavior\_units,{\rm{ }}behavior\_units} \right);$
\ENDFOR
\ENSURE ${k\_behavior\_units}$\; 
\end{algorithmic}
\end{algorithm}

For each sequence in the dataset, all subsequences of all possible lengths according to the min and max length parameters are visited. Algorithm 1 stores all candidates for a given behavior sequence with their associated quality measures (Line 8, which is detailed in Section \ref{s:measuring}). Once all behavior unit candidates have been assessed, first, they are sorted in order of quality, and self-similar behavior units are removed. Second, we merge these behavior units with the existing top k behavior units before processing the following behavior sequences. Last, we obtain the top k behavior units and discard all self-similar behavior units from the current sequences.

\subsubsection{Measuring the Quality of a Critical Behavior Unit}\label{s:measuring}
We denote the Euclidean distance between two subsequences ${W_S}$ and ${W_R}$ of length $l$ as
\begin{equation}
dist({W_S},{W_R}) = \sum\limits_{i = 1}^l {{{({s_i} - {r_i})}^2}} 
\end{equation}
The distance between a subsequence ${W_S}$  of length $l$  and behavior sequence ${I_i}$  is the minimum distance between $S$ and all normalized subsequences ${W_R}$  of ${I_i}$ i.e.
\begin{equation}
{d_{i,s}} = \mathop {\min }\limits_{R \in {W_{i,l}}} dist({W_S},{W_R})
\end{equation}
Therefore, all distances between a candidate behavior unit ${S_K}$  and behavior sequences $I=\{ {I_1},{I_2}, \ldots {I_n}\} $ are represented as a distance list ${D_k}$.

\begin{equation}
{D_k} =  < {d_{1,k}},{d_{2,k}},...{d_{n,k}} > 
\end{equation}
 
To further identify and extract critical behavior units, the shapelet algorithm \cite{ye2009time} is employed to determine the quality of a behavior unit. The original shapelet papers use information gain to determine the quality of a candidate shapelet \cite{ye2009time,lines2012shapelet}, because information gain is suitable for identifying how to obtain a partition of the data and can be applied to recursively divide the data. The original shapelet algorithm sorts the distance list ${D_k}$, and then evaluates the information gain for each possible split value. However, calculating distances between the behavior unit candidates and each behavior sequence is very time-consuming.

To solve this problem, we use the learning time-series shapelet model \cite{grabocka2014learning} to measure the quality of a critical behavior unit. First, a learning shapelet model is first trained. Second, we input each behavior unit candidate into the model and evaluate the quality according to the difference between the output of the model and the actual label.

We iteratively optimize the shapelet model by minimizing a classification loss function (which is shown in Equation \ref{eq_cross}) instead of searching among possible behavior units from all the behavior subsequences.

Given classifier weights  ${w}\in{R^{J+1}}$  (including bias) and a feature vector ${x_i} \in {R^{J}}$ , the linear prediction model is expressed as 
\begin{equation}
\hat y = \sum\limits_{j = 1}^J {{w_j}{x_{i,j}} + {w_0}}
\end{equation}
where ${{x_{i,j}}}$ is the distance between the $i$-th behavior sequence $I_i$ and the $j$-th shapelet $S_j$

The formulation jointly optimizes shapelets $S$ and classifier weights $w$ in
\begin{equation}
\mathop {{\rm{minimize}}}\limits_{S \in {R^{J \times L}},w \in {R^{J + 1}}} \;\sum\limits_{i = 1}^I {\mathcal{L} ({y_i},{{\hat y}_i}) + {\alpha  \over I}\sum\limits_{j = 1}^J {w_j^2} } 
\label{eq_cross}
\end{equation} 
where $\alpha  \ge 0$ is a regularization parameter. Considering class labels $Y = \left\{ {1,0} \right\}$, the loss function $\mathcal{L}$ is 
\begin{equation}
\mathcal{L} ({y_i},{\hat y_i}) =  - {y_i}\ln (\sigma ({\hat y_i})) - (1 - {y_i})\ln (1 - \sigma ({\hat y_i}))
\end{equation}  
and the sigmoid function $\sigma $ is
\begin{equation}
\sigma ({\hat y_i}) = {(1 + {e^{ - {{\hat y}_i}}})^{ - 1}}
\end{equation}   

Then, the shapelets $S = \{ {S_1},{S_2}, \ldots {S_J}\} $ and classifier weights $w$ can be learned to minimize the classification objective and reduce generalization errors without compromising shapelet interpretability.

We input each behavior unit candidate ${W_i}$ into the model and evaluate these qualities according to the difference $\xi $ between the output of the model $\hat y({W_i})$  and the actual label ${y_{actual}}$  of ${W_i}$,
\begin{equation}
\xi  = {y_{actual}} - \hat y({W_i}) = {y_{actual}} - \sum\limits_{j = 1}^J {{w_j}{d_{{W_i},{S_j}}} - {w_0}}
\end{equation}   
where ${d_{{W_i},{S_j}}}$ is the distance between the behavior unit candidate ${W_i}$ and the $j$-th shapelet $S_j$. We use $\xi$ to evaluate the quality of candidates. If $\xi$ is close to 0, the quality of the behavior unit is high.
 
\subsection{Behavior Unit Extraction and Representation}

\subsubsection{Behavior Unit Extraction}

By extracting the critical behavior units, we exclude unrelated behavior sequences from the perturbed sequences and improve the robustness against obfuscation attacks. The token-level longest common sequence (LCS) \cite{bergroth2000survey} is employed for the extraction. We assume that ${I_x} = ({x_1},{x_2},...,{x_n})$  and  ${I_y} = ({y_1},{y_2},...,{y_m})$ are a behavior unit and behavior sequence of length $n$ and $m$, respectively (${\rm{n <  < m}}$). Their LCS is represented by ${\rm{LCS}}({I_x},{I_y}) = {\rm{M}}(n,m)$ , where   :
\begin{equation}
{\rm{M}}(i,j) = \left\{ \begin{array}{l}
1 + {\rm{M(}}i - 1,j - 1){\rm{  \;\;\,    }};{\rm{ }}{x_i} = {y_j}{\rm{\;}}i,j > 0\\
{\rm{Max}}\left\{ \begin{array}{l}
{\rm{M}}(i - 1,j)\\
{\rm{M}}(i,j - 1)
\end{array} \right.{\rm{ }};{\rm{ }}{x_i} \ne {y_j}{\rm{\; }}i,j > 0\\
0{\rm{\;\;\;\;\;\;\;\;\;\;\;\;\;\;\;\;\;\;\;\;\;\;\;\;\;\;\;\;\;\;}};{\rm{ }}i = 0,{\rm{ }}or{\rm{\; }}j = 0
\end{array} \right.
\end{equation}

When the length of ${\rm{M}}(n,m)$  is $n$, the behavior unit ${S_x}$ is completely contained in the behavior sequence ${S_y}$. The LCS-based behavior unit extraction process is shown in Step 3 of Figure \ref{fig_frame}. For instance, we can obtain the behavior unit set $\alpha $= $\{$D→C→H, B→G→A,D→F$\}$ by behavior unit identification. Given a behavior sequence $\beta $ = $\{$D→E→B→G→E→C→H→A$\}$, we match the identified behavior units from $\alpha $ on the behavior sequence $\beta $ and retain only the behavior units that are completely contained in the behavior sequence. Then, we can obtain the extracted behavior sequence $\gamma $=$\{$D→B→G→C→H→A$\}$.

\subsubsection{Behavior Sequence Representation}

\begin{figure*}
\centering
\includegraphics[scale=0.5]{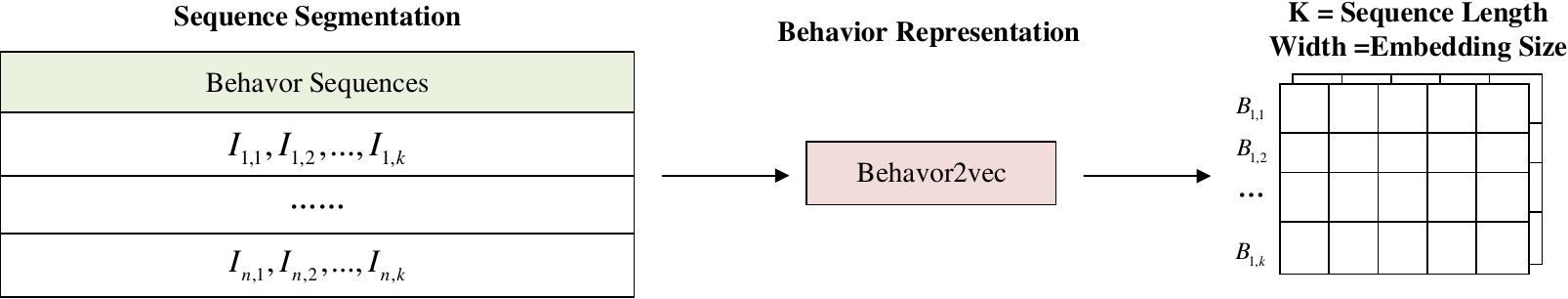}
\caption{\centering{Behavior sequence representation.}}
\label{fig_b2v}
\end{figure*}

To characterize behaviors by similarity of semantic and context, we employ the word2vec \cite{al2019use} model to represent the behaviors as vectors, as is shown in Figure \ref{fig_b2v}. The word2vec model can make the embeddings of different behaviors with similar semantics and contexts closer in the representation space. We leverage the skip-gram method to train the word2vec based on the raw input behavior sequences in the dataset.

Let behavior sequences ${I} = ({I_1},{I_2},...,{I_n})$, the representation of the behavior sequence is 
\begin{equation}
{B_e} = Behavior2Vec(I){\rm{ ,\;  }}{B_e} \in {R^{K \times Q}}
\end{equation}

\subsection{Feature Extraction and Behavior Classification}
After the previous steps, we can obtain the new behavior sequences from the perturbed sequences based on the identified behavior units.

Since sequential deep learning models (e.g., LSTM and RNN) have a good ability to learn sequential features, abnormal behavior sequences can usually be detected by these models. However, we argue that existing deep learning models are not sufficient to represent different granularity levels of behavior semantics in behavior sequences. A behavior unit contains the collection of strongly related behaviors, which represents the key classification features of related behaviors. Therefore, the features of behavior units are necessary for behavior analysis.

\begin{figure}
\centering
\includegraphics[scale=0.5]{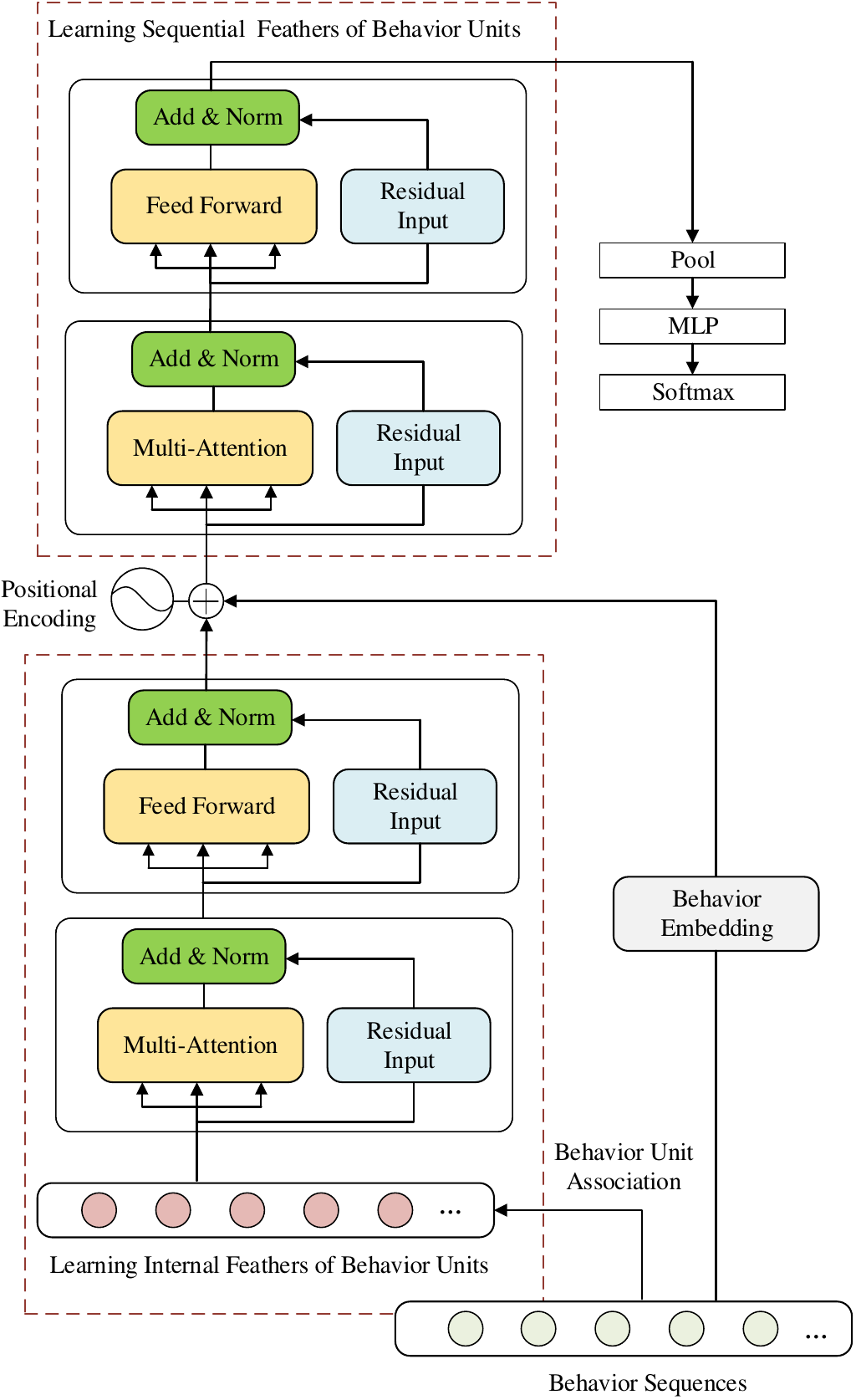}
\caption{\centering{The architecture of the proposed classification model.}}
\label{fig_transform}
\end{figure}

Based on this insight, we design a multilevel transformer-based model to learn the joint features within and between behavior units, thereby improving the accuracy and antiattack ability of the detection method. The overall architecture of our classification model is shown in Figure \ref{fig_transform}.

\textbf{1) Learning Internal Features of Behavior Units}

\begin{figure*}
\centering
\includegraphics[scale=0.5]{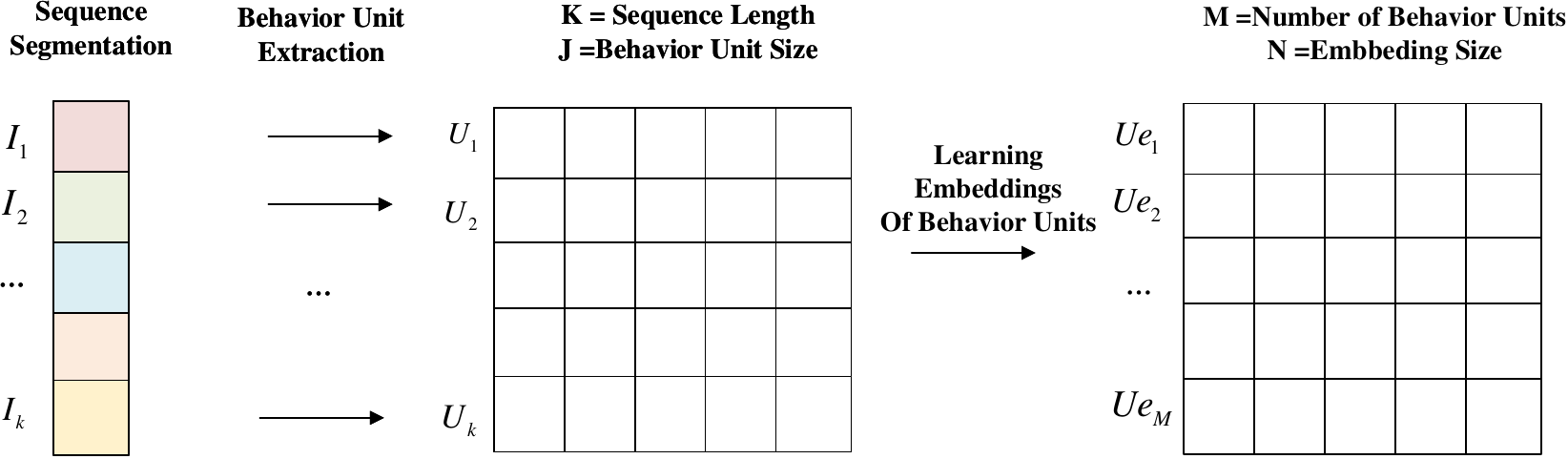}
\caption{\centering{Learning the internal features of behavior units.}}
\label{fig_internal}
\end{figure*}

Since a behavior unit can represent the critical classification features of related behaviors, we provide a transformer encoder block to learn the embedded representation of different behavior units. The learning process is shown in Figure \ref{fig_internal}.

We only use the transformer encoder for our anomaly detection model, because it is designated for learning the features at different levels of behaviors. This step is achieved by employing a multi-head self-attention mechanism, which allows the model to selectively focus on different parts of the input sequence at each level. The transformer encoder is known to be highly effective for sequence modeling tasks, as it is based on the self-attention mechanism that can establish long-range dependencies among different positions. This kind of dependency can help the model better capture the structure and features of the data, and thus better distinguish between normal data and anomalous data. 

We did not use the decoder part of the Transformer model, because our anomaly detection problem is focused on learning the feature representations of input sequences and identifying anomalies in the input sequences rather than generating a new sequence based on a given input sequence. Therefore, the decoder part, which is responsible for generating the output sequence based on the encoded input sequence, is not relevant to our task. 

Specifically, we associate each behavior ${I_i}$  with its corresponding behavior unit ${U_i}$. Assuming that the behavior unit ${U_i}$ contains behaviors of $J$ length, where ${U_i} = \{ {B_{i,1}},{B_{i,2}},...{B_{i,J}}\} $, we input each behavior unit into the transformer encoder block, and use the last layer of the transformer encoder as the embedded representation of the behavior unit because it contains the richest information after multiple layers of self-attention. For the n layers of the transformer encoder block, the representations ${U_e}$  from the $n$-th layer are denoted as:

\begin{equation}
Ue = \phi _{transformer}^{(n)}(I) = \{ h_1^{(n)},h_2^{(n)},...,h_J^{(n)}\} 
\end{equation}

\textbf{2) Learning Sequential Features of Behavior Units}

We use other transformer encoder blocks to learn the sequential features of behavior units. 

We concatenate the embeddings of behavior unit representations with   the embeddings of behavior representations as input and feed the joint embeddings into the transformer encoder blocks to learn contextual features of the behavior and behavior unit sequences:
\begin{equation}
\;input{\rm{ = concentrate (}}Be{\rm{, }}Ue{\rm{)}}
\end{equation} 

\textbf{3) Behavior Classification }

The outputs of the transformer encoder blocks are sent to a pooling layer and an MLP classification layer. The softmax layer is employed to classify the behaviors as normal or abnormal. 

\section{Evaluation}\label{s:evaluation}
In this section, first, we evaluate the detection performance and adversarial robustness ability of the model and make a comparison with many baseline models \cite{du2017deeplog,nedelkoski2020self,ludetecting,farzad2020unsupervised}. Second, we conduct ablation studies to analyze the contribution of each defense process of our model. Last, we compared our model with other defense approaches. 

\subsection{Environment}
The following experiments are performed on a Windows 10 operating system, powered by AMD Ryzen 7 5800 8-Core Processor 3.40 GHz, NVIDIA 3060 and 32.0 GB of RAM. Keras version 2.7 is employed to implement our model.

\subsection{Dataset and Sequence Generation}

We use the AndroCT dataset\cite{li2021androct}, which is a dataset for Android malware detection that contains more than 35,974 Android applications collected from 2010 through 2019, including malicious and benign applications. Behavior data consist of API invocation sequences. In our experiments, we use the latest version (i.e., 2019) of behavior logs. 

We apply a fixed window grouping with different sizes of log records to generate behavior sequences on this dataset. We label API sequences of malware "positive" and those used for benign software "negative".

We use a method similar to \cite{rosenberg2020query} to implement a state-of-the-art adversarial attack focused on malware, as shown in Algorithm \ref{alg2}. Instead of intercepting benign fragments directly from the behavior execution sequences, we employ seqGAN \cite{yu2017seqgan} to generate irrelevant behaviors and insert them into the original sequences. The intercepted fragments are more likely to be detected as ``adversarial signatures" with obvious insertion marks. 

\begin{algorithm} 
\caption{Adversarial Sequence Generation} 
\label{alg2}
\begin{algorithmic}[1]
\REQUIRE $x$ (malicious sequence to perturb, of length $l$), 
 
 \noindent \,\;\;\;\;\;$n$ (number of adversarial sliding window),

 \noindent \;\;\;\; $B$ (max injection rate of generated benign fragments)

\FOR{each sliding window $ w_j $ of $n$  in $x$:}
\WHILE { curr\_injection rate in $x$ $<$ $B$ }
\STATE   Randomly select a behavior’s position $i$ in $ w_j $
\STATE   Insert a new adversarial fragment

at position $i$ of $ w_j $, $i \in \left\{ {1,2,...,n} \right\}$
\ENDWHILE
\ENDFOR
\ENSURE perturbed $x$
\end{algorithmic}
\end{algorithm}

\subsection{Evaluation Method}
The confusion matrix is applied to evaluate the performance of our proposed model. Let TP represent the number of sequences that are correctly predicted as positive, TN denote the number of sequences that are correctly classified as negative, FN denote the number of traces that are positive but are incorrectly predicted as negative, and FP indicate the number of traces that are negative but are predicted as positive. We measure our model in terms of accuracy, precision, recall, and F1 score to assess our detection performance and make comparisons. In our experiment, we use fixed windows to divide behavior logs into behavior sequences, and the goal of our classification is to determine whether the behavior sequence is malicious. 

\begin{equation}
Precision = {{TP} \over {TP + FP}}
\end{equation}

\begin{equation}
 Recall = {{TP} \over {TP + FN}}
\end{equation}

\begin{equation}
 F1 - Score = {{2*Precision*Recall} \over {Precision + Recall}} 
\end{equation}

\subsection{Baselines}
\subsubsection{Log Behavior Analysis Approaches}

The proposed model is compared with several baseline models. All the baselines focus on log-based behavior analysis approaches, which are open-source 
and based on sequential deep learning models. These approaches are presented as follows:

\textbf{1) Min Du et al. (2017) \cite{du2017deeplog}:} This approach is named Deeplog, which adopts LSTM networks to learn the behavior patterns of logs. The input is a one-hot vector for each behavior pattern. 

\textbf{2) Sasho Nedelkoski et al. (2020) \cite{nedelkoski2020self}:} They adopt a Transformer encoder with a multi-head self-attention mechanism, which learns context information from behavior logs in the form of log vector representations (i.e., embeddings).

\textbf{3) Siyang Lu et al. (2018) \cite{ludetecting}:} This approach performs behavior log anomaly detection by leveraging convolutional neural networks (CNN), which can explore the latent complex relationships in behavior logs.

\textbf{4) Amir Farzad et al. (2020) \cite{farzad2020unsupervised}:} They propose an unsupervised model for behavior log anomaly detection that employs two deep autoencoder networks for anomaly detection.

\subsubsection{Defense Approaches of Adversarial Attacks}

We compare our approach with several typical defense approaches \cite{rosenberg2019defense} against behavior log adversarial attacks. 

\textbf{1) Sequence Squeezing \cite{rosenberg2021sequence}:} Sequence squeezing reduces the search space available to an adversary by merging similar semantic features into a single representative feature. For instance, the syscalls “sys$\_$read” and “sys$\_$read” can be merged into a system operation behavior of “reading file”. To implement this approach, we use word2vec to represent the syscall names as word embeddings and cluster similar syscalls with the same semantics (by using Euclidean distance). Therefore, different merged groups can represent different operation behaviors. 

\textbf{2) Defense Sequence-GAN \cite{rosenberg2019defense}:} This approach defends adversarial attacks by training a GAN to learn the distribution of unperturbed behavior sequences. Therefore, the Defense Sequence-GAN can generate approximate samples that meet the unperturbed sample distribution. We employ SeqGAN to implement this approach. We train a benign SeqGAN and malicious SeqGAN using the unperturbed dataset and generate benign samples and malicious samples, respectively. When an input sequence emerges, we choose the generated unperturbed sequence nearest the perturbed sequence (calculated by Euclidean distance) and feed it to the classifier.

\textbf{3) Adversarial Learning \cite{szegedy2013intriguing}:} Adversarial learning adds adversarial samples to the training set, which can make the classifier learn the distribution of adversarial samples, thereby defending against adversarial attacks. We generate malicious adversarial samples according to Algorithm \ref{alg2} and add them to our training dataset. In addition, we label these samples abnormal. Then, we train the classifier using the dataset and analyze the detection performance.

\textbf{4) GuardOL \cite{das2015semantics}:} This approach constructs a collection of critical events, and the feature vector of software is generated by extracting the frequency of each event to perform classification based on the MLP model. Since the dataset used by this approach comprises syscall logs, we directly use the critical behaviors identified by our approach in our dataset to extract the frequency of each software and then use the MLP as the classifier.

\subsection{Results}
\subsubsection{Robustness to Adversarial Attacks }

To prove the ability and robustness of our proposed model, we design the method of adversarial attack focused on behavior sequences and conduct comparative experiments among the proposed model and other models in \cite{du2017deeplog,nedelkoski2020self,ludetecting,farzad2020unsupervised} by taking into account the adversarial attacks.

Specifically, we employ seqGAN \cite{yu2017seqgan}, a contextualized generation model, to generate benign fragments. To prevent damage to the functionality of the malware, we follow the idea of generating adversarial sequence samples mentioned in \cite{rosenberg2020query}. We insert these adversarial samples into the original sequences. 

\begin{table*}[ht]%
\renewcommand\arraystretch{2}
\centering 
\caption{A behavior sequence example.}
\label{tab_sample}
\begin{tabular}{c|c} \hline 
Original Sequence & size booleanValue  hashCode equals valueOf  \\ \hline
Adversarial Sequence &	size booleanValue \framebox{\textbf{readLine}} \framebox{ \textbf{replace}} hashCode  \framebox{\textbf{gethtml}} equals valueOf \\ \hline
\end{tabular}
\end{table*}

\begin{table*}[ht]
\renewcommand\arraystretch{1.2}
\centering
\caption{The robustness to adversarial attacks of different methods.}
\label{tab_rob}
\begin{tabular}{cp{4.19em}ccccccccc}
    \toprule
    \multirow{2}[3]{*}{Method} & \multirow{2}[3]{*}{Metrics} & \multicolumn{9}{p{30em}}{~~~~~~~~~~~~~~~~~~~~~~~~~~~~~~~~~~~~~~~~~~~~~Injection Rate} \\
\cmidrule{3-11}    & \multicolumn{1}{c}{} & 0\%   & 5\% & 10\%  & 15\%  & 20\%  & 25\%  & 30\% & 35\% & 40\%\\
    \midrule
    \multirow{2}[3]{*}{LSTM-based Model \cite{du2017deeplog}} & ~~~~R & 0.992 & 0.931 & 0.885 & 0.828 & 0.797 & 0.753 & 0.725 & 0.665 & 0.607 \\
          & ~~~~P     & 0.982 & 0.993 & 0.991 & 0.991 & 0.989 & 0.989 & 0.989 & 0.991 & 0.990 \\
          & ~~~~F1    & \textbf{0.987} & \textbf{0.961} & \textbf{0.935} & \textbf{0.902} & \textbf{0.883} & \textbf{0.855} & \textbf{0.837} & \textbf{0.796} & \textbf{0.753} \\
    \midrule
    \multirow{2}[3]{*}{Transformer-based Model \cite{nedelkoski2020self}} & ~~~~R & 0.985 & 0.945 & 0.904 & 0.865 & 0.825 & 0.801 & 0.761 & 0.733 & 0.702 \\
          & ~~~~P     & 0.984 & 0.992 & 0.992 & 0.990  & 0.992 & 0.988 & 0.990  & 0.992 & 0.992 \\
          & ~~~~F1   & \textbf{0.985} & \textbf{0.968} & \textbf{0.946} & \textbf{0.923} & \textbf{0.901} & \textbf{0.885} & \textbf{0.861} & \textbf{0.843} & \textbf{0.822} \\
    \midrule
    \multirow{2}[3]{*}{CNN-based Model\cite{ludetecting}} & ~~~~R  & 0.978 & 0.929 & 0.862 & 0.842 & 0.778 & 0.742 & 0.706 & 0.667 & 0.633 \\
          & ~~~~P     & 0.990  & 0.995 & 0.995 & 0.989 & 0.989 & 0.995 & 0.993 & 0.994 & 0.994 \\
          & ~~~~F1      & \textbf{0.984} & \textbf{0.961} & \textbf{0.924} & \textbf{0.91} & \textbf{0.871} & \textbf{0.85} & \textbf{0.825} & \textbf{0.806} & \textbf{0.774} \\
    \midrule
    \multirow{2}[3]{*}{Autoencoder-based Model \cite{farzad2020unsupervised}} & ~~~~R     & 0.988 & 0.962 & 0.921 & 0.872 & 0.857 & 0.808 & 0.775 & 0.733 & 0.724 \\
          & ~~~~P     & 0.824 & 0.814 & 0.810  & 0.807 & 0.806 & 0.800   & 0.799 & 0.795 & 0.789 \\
          & ~~~~F1      & \textbf{0.899} & \textbf{0.882} & \textbf{0.862} & \textbf{0.839} & \textbf{0.831} & \textbf{0.804} & \textbf{0.787} & \textbf{0.763} & \textbf{0.755} \\
    \midrule
    \multirow{2}[3]{*}{Our Model}  & ~~~~R & 0.968 & 0.969 & 0.949 & 0.947 & 0.935 & 0.954 & 0.911 & 0.932 & 0.902 \\
          & ~~~~P     & 0.999 & 0.974 & 0.987 & 0.969 & 0.971 & 0.944 & 0.983 & 0.938 & 0.944 \\
          & ~~~~F1    & \textbf{0.983} & \textbf{0.971} & \textbf{0.968} & \textbf{0.958} & \textbf{0.953} & \textbf{0.949} & \textbf{0.945} & \textbf{0.935} & \textbf{0.923} \\
    \bottomrule
    \end{tabular}%
\end{table*}

We divided the dataset into two equal parts: DataSet1 and DataSet2. Then, we use normal sequences of DataSet1 to train the SeqGAN model and generate benign fragments, while the victim models are trained on DataSet2. We generate adversarial samples according to Algorithm \ref{alg2}. These samples are fed into the trained model. Table \ref{tab_sample} shows an adversarial example, and the inserted behaviors are highlighted. 

Table \ref{tab_rob} shows the detection rates on Dataset2 when the injection rate percentage increases from ${\rm{0\% }}$ to ${\rm{40\% }}$. The detection rates of the models that suffer adversarial attacks have been reduced. Specifically, the F1 score of Deeplog drops from ${\rm{98.7\% }}$ to ${\rm{75.3\% }}$ when the injection rate is ${\rm{40\% }}$. These results show that the context characteristics of malicious sequences can obviously be changed by inserting the generated benign fragments.

In contrast, our model is effective in defending adversarial samples. As shown in Table \ref{tab_rob}, the F1 score drops only 6\% in the worst case. The results show that our proposed model, based on critical behavior unit learning, is more robust to defend against adversarial attacks than other approaches. The robustness comes from two aspects. Specifically, in the testing process, we only focus on learning critical behavior units, and the inserted fragments in the adversarial samples are filtered out. In addition, our model learns the overall semantics of each critical behavior unit and the contextual relationships among behavior units. Therefore, the intentions of malicious behaviors can be extracted and learned.

\subsubsection{Ablation Studies}

In the process of feature extraction and behavior classification, the features of behavior units are learned. To assess the contribution of behavior unit features, we remove behavior unit features from the model and then evaluate the performance of the remaining model. 

It shows that the lack of behavior unit feature learning causes performance degradation. According to the results in Table \ref{tab_abla2}, the model without behavior unit features has worse performance, as the F1 scores of the model decrease by ${\rm{3.1\% }}$ and ${\rm{0.8\% }}$ when the injection rates are ${\rm{20\% }}$ and ${\rm{40\% }}$, respectively. Therefore, behavior unit features have an important impact on the proposed model. This finding indicates that the behavior unit features can perceive a wider range of behaviors and improve the anomaly detection and anti-obfuscation ability.

\begin{table}
\renewcommand\arraystretch{1.2}
\centering
\caption{The ablation experimental results with different injection rates.}
\label{tab_abla2}
\begin{tabular}{cp{4em}cccc}
    \toprule
    \multirow{2}[1]{6em}{~~Method} & \multirow{2}[1]{6em}{Metrics} & \multicolumn{2}{p{8em}}{~~~~Injection Rate} \\
\cmidrule{3-4}    & \multicolumn{1}{c}{} & 20\%  & 40\%\\
    \midrule
    \multirow{2}[1]{8em}{Without behavior unit feathers } & ~~~~R & 0.921 & 0.876  \\
          & ~~~~P     & 0.923 & 0.956  \\
          & ~~~~F1    & \textbf{0.922} & \textbf{0.915}  \\
    \midrule
    \multirow{2}[1]{8em}{With behavior unit feathers } & ~~~~R & 0.935 & 0.902  \\
          & ~~~~P     & 0.971 & 0.944  \\
          & ~~~~F1   & \textbf{0.953} & \textbf{0.923} \\
    \bottomrule
    \end{tabular}%
\end{table}

\begin{table*}
\renewcommand\arraystretch{1.2}
\centering
\caption{The results of improved models with behavior unit extraction and identification.}
\label{tab_abla}
\begin{tabular}{cp{4.19em}cccccccc}
    \toprule
    \multirow{2}[3]{*}{Method} & \multirow{2}[3]{*}{Metrics} & \multicolumn{8}{p{27em}}{~~~~~~~~~~~~~~~~~~~~~~~~~~~~~~~~~~~~~Injection Rate} \\
\cmidrule{3-10}    & \multicolumn{1}{c}{} & 5\% & 10\%  & 15\%  & 20\%  & 25\%  & 30\% & 35\% & 40\%\\
    \midrule
    \multirow{2}[3]{*}{LSTM-based Model \cite{du2017deeplog}} & ~~~~R & 0.986 & 0.963 & 0.953 & 0.942 & 0.926 & 0.863 & 0.842 & 0.813 \\
          & ~~~~P      & 0.936 & 0.938 & 0.936 & 0.935 & 0.937 & 0.935 & 0.936 & 0.936 \\
          & ~~~~F1    & \textbf{0.961} & \textbf{0.950} & \textbf{0.944} & \textbf{0.938} & \textbf{0.931} & \textbf{0.897} & \textbf{0.886} & \textbf{0.870} \\
    \midrule
    \multirow{2}[3]{*}{Transformer-based Model \cite{nedelkoski2020self}} & ~~~~R & 0.994 & 0.973 & 0.963 & 0.960 & 0.923 & 0.871 & 0.860 & 0.854 \\
          & ~~~~P     & 0.945 & 0.936 & 0.940 & 0.935 & 0.935 & 0.937 & 0.937 & 0.933 \\
          & ~~~~F1   & \textbf{0.968} & \textbf{0.954} & \textbf{0.951} & \textbf{0.947} & \textbf{0.929} & \textbf{0.902} & \textbf{0.897} & \textbf{0.892} \\
    \midrule
    \multirow{2}[3]{*}{CNN-based Model\cite{ludetecting}} & ~~~~R   & 0.986 & 0.947 & 0.933 & 0.923 & 0.889 & 0.854 & 0.765 & 0.744 \\
          & ~~~~P      & 0.942 & 0.941 & 0.942 & 0.938 & 0.937 & 0.939 & 0.941 & 0.937 \\
          & ~~~~F1      & \textbf{0.963} & \textbf{0.944} & \textbf{0.937} & \textbf{0.930} & \textbf{0.912} & \textbf{0.894} & \textbf{0.844} & \textbf{0.829} \\
    \midrule
    \multirow{2}[3]{*}{Autoencoder-based Model \cite{farzad2020unsupervised}} & ~~~~R      & 0.973 & 0.933 & 0.907 & 0.897 & 0.864 & 0.841 & 0.812 & 0.767 \\
          & ~~~~P     & 0.828 & 0.824 & 0.829 & 0.817 & 0.824 & 0.831 & 0.831 & 0.825 \\
          & ~~~~F1      & \textbf{0.894} & \textbf{0.875} & \textbf{0.866} & \textbf{0.855} & \textbf{0.843} & \textbf{0.836} & \textbf{0.821} & \textbf{0.795} \\

    \bottomrule
    \end{tabular}%
\end{table*}

\begin{table*}
\renewcommand\arraystretch{1.2}
\centering
\caption{Comparison with other defense methods.}
\label{tab_other}
\begin{tabular}{cp{4.19em}cccccccc}
    \toprule
    \multirow{2}[3]{*}{Method} & \multirow{2}[3]{*}{Metrics} & \multicolumn{8}{p{27em}}{~~~~~~~~~~~~~~~~~~~~~~~~~~~~~~~~~~~~~~~Injection Rate} \\
\cmidrule{3-10}    & \multicolumn{1}{c}{} & 5\% & 10\%  & 15\%  & 20\%  & 25\%  & 30\% & 35\% & 40\%\\
    \midrule
    \multirow{2}[3]{*}{Sequence Squeezing \cite{rosenberg2021sequence}} & ~~~~R & 0.943 & 0.902 & 0.859 & 0.819 & 0.789 & 0.758 & 0.730  & 0.719 \\
          & ~~~~P      & 0.985 & 0.982 & 0.985 & 0.985 & 0.985 & 0.985 & 0.985 & 0.971 \\
          & ~~~~F1     & \textbf{0.964} & \textbf{0.940} & \textbf{0.918} & \textbf{0.895} & \textbf{0.876} & \textbf{0.857} & \textbf{0.838} & \textbf{0.826} \\
    \midrule
    \multirow{2}[3]{*}{Defense Sequence-GAN \cite{rosenberg2019defense}} & ~~~~R & 0.907 & 0.873 & 0.841 & 0.815 & 0.794 & 0.771 & 0.752 & 0.717 \\
          & ~~~~P      & 0.984 & 0.979 & 0.978 & 0.983 & 0.972 & 0.967 & 0.970  & 0.975 \\
          & ~~~~F1   & \textbf{0.944} & \textbf{0.923} & \textbf{0.905} & \textbf{0.891} & \textbf{0.874} & \textbf{0.859} & \textbf{0.847} & \textbf{0.826} \\
    \midrule
    \multirow{2}[3]{*}{Adversarial Learning\cite{szegedy2013intriguing}} & ~~~~R   & 0.967 & 0.937 & 0.894 & 0.871 & 0.844 & 0.828 & 0.806 & 0.771 \\
          & ~~~~P      & 0.961 & 0.969 & 0.961 & 0.962 & 0.962 & 0.955 & 0.964 & 0.947 \\
          & ~~~~F1      & \textbf{0.964} & \textbf{0.953} & \textbf{0.927} & \textbf{0.914} & \textbf{0.899} & \textbf{0.887} & \textbf{0.878} & \textbf{0.850} \\
    \midrule
    \multirow{2}[3]{*}{GuardOL  \cite{das2015semantics}} & ~~~~R      & 0.935 & 0.931 & 0.925 & 0.915 & 0.910  & 0.905 & 0.901 & 0.895 \\
          & ~~~~P      & 0.930  & 0.930  & 0.930  & 0.929 & 0.929 & 0.928 & 0.928 & 0.927 \\
          & ~~~~F1       & \textbf{0.933} & \textbf{0.93} & \textbf{0.927} & \textbf{0.922} & \textbf{0.919} & \textbf{0.916} & \textbf{0.914} & \textbf{0.911} \\
    \midrule
    \multirow{2}[3]{*}{Our Approach } & ~~~~R      & 0.969 & 0.949 & 0.947 & 0.935 & 0.954 & 0.911 & 0.932 & 0.902 \\
          & ~~~~P     & 0.974 & 0.987 & 0.969 & 0.971 & 0.944 & 0.983 & 0.938 & 0.944 \\
          & ~~~~F1     & \textbf{0.971} & \textbf{0.968} & \textbf{0.958} & \textbf{0.953} & \textbf{0.949} & \textbf{0.945} & \textbf{0.935} & \textbf{0.923} \\
    \bottomrule
    \end{tabular}%
\end{table*}

\begin{figure}
    \begin{minipage}[t]{0.5\linewidth}
        \centering
        \includegraphics[width=\textwidth]{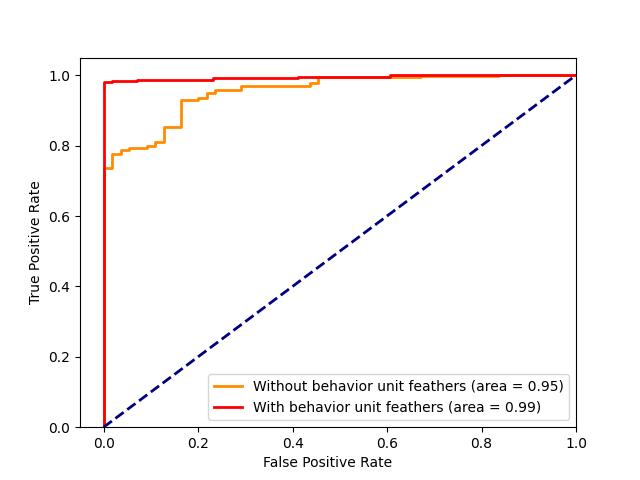}
        \centerline{(a) Injection rate of 20$\%$  }
    \end{minipage}%
    \begin{minipage}[t]{0.5\linewidth}
        \centering
        \includegraphics[width=\textwidth]{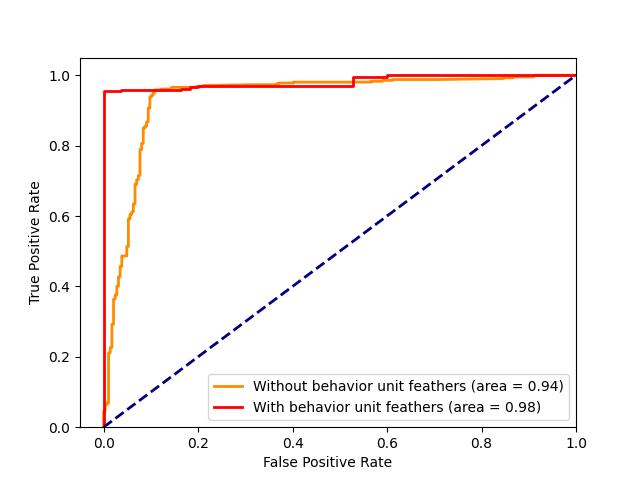}
        \centerline{(b) Injection rate of 40$\%$ }
    \end{minipage}
    \caption{ROC curves of the proposed model with different injection rates. }
    \label{f:ROC-ablation}
\end{figure}

Figure \ref{f:ROC-ablation} shows the ROC curves of the proposed model and that the AUC score decreases after the removal of behavior unit features. Thus, behavior unit features have a good effect on malware detection, which enables the model to adapt to complex detection situations.

Additionally, to verify the effectiveness of behavior unit extraction and identification, we input the sequences obtained by these two processes into the baseline models and evaluate whether the performance of these baseline models is improved. As shown in Table \ref{tab_abla}, the detection performance of all models has been improved. Overall, with the increasing injection ratio of benign fragments, these improved baseline models can maintain higher F1 scores than the original models, which implies that behavior unit extraction and identification can improve the ability to resist adversarial sample attacks.

\subsubsection{Comparison with Other Defense Methods}

We apply the above-mentioned adversarial defense methods as baselines for comparison with the same datasets and in the same experimental setting. The results are shown in Table \ref{tab_other}. Overall, our proposed approach achieves the highest performance. As the injection rate of adversarial attacks increases, the detection ability of our model maintains the best performance.

The Defense Sequence-GAN method cannot handle adversarial samples with long sequence injection, so its performance is not good. In addition, this method needs to identify the samples most similar to the input among a large number of generated samples, which is very time-consuming.

Sequence Squeezing mainly defends against critical behavior obfuscation attacks, which replace well-known malicious behaviors with less well-known behaviors with similar functionality. However, this approach cannot be applied to identifying perturbations with benign behaviors.

As the proportion of malicious sample injection increases, the F1 score of adversarial learning drops by 11.4\%, which is much higher than our approach. That is because the generated perturbed samples are very similar to normal samples and reduce the detection ability of the model.

The performance of GuardOL is better than that of other baselines, which indicates that identifying critical behaviors is important for the robustness of the model. Since this approach cannot analyze the sequential features of behaviors, its performance is worse than that of our model.

\subsubsection{Sustainability Analysis}

Our approach is related to sustainability analysis\cite{cai2020embracing,cai2020assessing}. Although the use of artificially generated adversarial samples for evaluation purposes is common practice, it may not accurately represent real-world adversarial attack scenarios. Therefore, it is crucial to incorporate real-world adversarial malware samples. Software may change its behaviors over time as it evolves. We can use the behavior logs of the software for a certain year as unperturbed data and then use the logs of later versions as perturbed behavior data.


In addition, we compare our approach with other baselines. The targets analyzed by these methods are mainly complete programs, whereas the targets we analyzed in the previous tests were independent behavioral sequences. In order to compare with baselines at the software level, we organize the behavior sequences generated by each program into groups. If any behavior sequence within a group is identified as abnormal, it is classified as malware. The baselines are as follows.

\textbf{1) DroidSpan \cite{cai2020assessing}:} It is a novel behavior profile to effectively capture the distribution of sensitive access within Android apps. Through a longitudinal analysis, consistent distinctions between benign apps and malware are identified, persisting over a duration of seven years despite the evolving nature of both app types.


\textbf{2) DroidEvolver \cite{xu2019droidevolver}:} Since different models have different sensitivities to software behavior evolution, this method establishes a model pool composed of multiple machine learning models and performs software classification through multi-model voting. 

\textbf{3) MamaDroid \cite{onwuzurike2019mamadroid}:} It utilizes Markov chain modeling of API call sequences for Android malware detection. It offers three modes of operation for abstracting API calls at different granularities: families, packages, or classes. It achieves high accuracy in detecting unknown malware samples.

We train our model and the baselines on the AndroCT dataset of years 2013 and 2014 and evaluate their performance on data from the subsequent years. For a fair comparison, we use the same samples in the AndroCT dataset as DroidSpan \cite{cai2020assessing} to test all baselines and our approach. The numbers of benign and malicious samples in the test are shown in Table \ref{tab_6}.

\begin{table}
\renewcommand\arraystretch{1.2}
\centering
\caption{The number of samples from each year used on the AndroCT dataset.}  
\label{tab_6}
%
\begin{tabular}{ccc} \hline 
Year &  Number of Benign APPs & Number of Malware\\\hline
2013    & 1568 & 1139 \\
2014     & 2953  & 1337\\
2015     & 1178   & 1451\\
2016     & 1370  & 1769\\
2017     & 1612  & 1934 \\  \hline
\end{tabular}
\end{table}

           

\begin{figure*}
    \begin{minipage}[t]{0.5\linewidth}
        \centering
        \includegraphics[width=\textwidth]{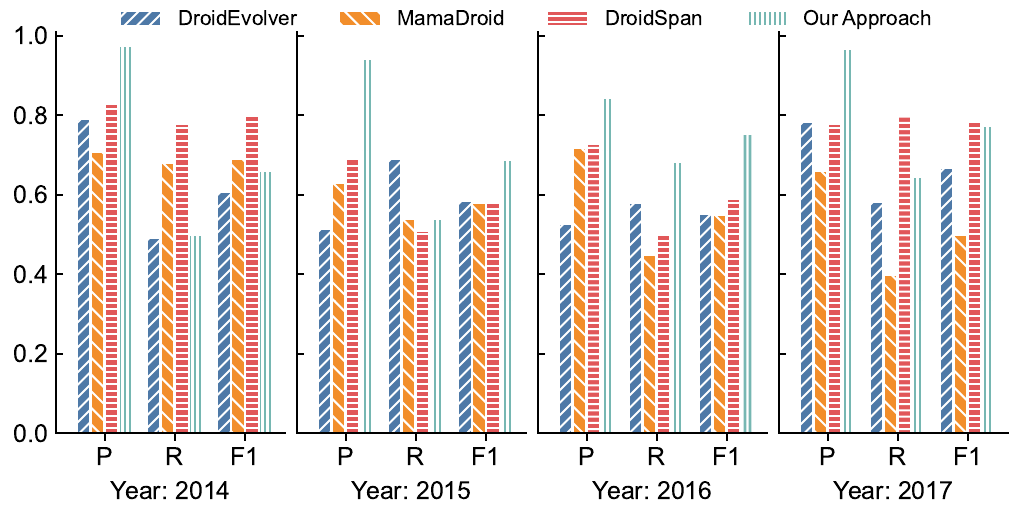}
        \centerline{(a) Trained on the 2013 dataset  }
    \end{minipage}%
    \begin{minipage}[t]{0.5\linewidth}
        \centering
        \includegraphics[width=\textwidth]{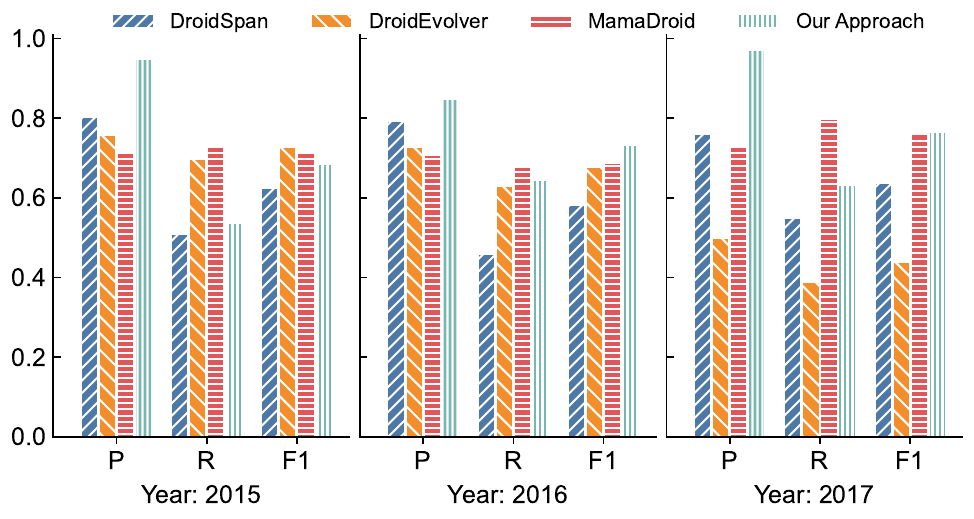}
        \centerline{(b) Trained on the 2014 dataset }
    \end{minipage}
    \caption{Performance of different models trained on the data of 2013 and 2014 and tested on the data of subsequent years.}
    \label{fig_7}
\end{figure*}

\begin{figure*}
\centering
\includegraphics[scale=0.55]{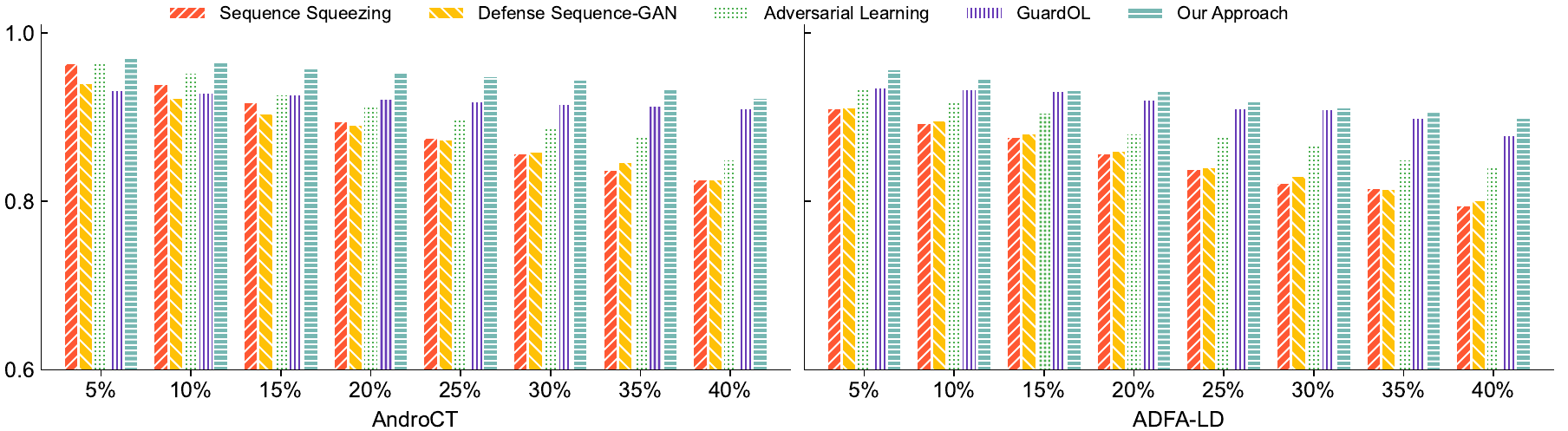}
\caption{The F1 scores of our approach and other four adversarial defense approaches under different attack injection rates on two datasets.}
\label{fig_8}
\end{figure*}

The experimental results are shown in Figure \ref{fig_7}. Among the baselines, DroidSpan achieves the best performance. For instance, it gets the highest average F1 score of 0.691 across the years 2014-2017 among the baselines when trained on the data of 2013. When it is trained on the data of 2014 and tested on the data of 2015-2017, the average F1 score is 0.723, which is also the highest among the baselines. The results demonstrate its effectiveness in sustainable malware detection of Android APPs. DroidEvolver achieves average F1 scores of 0.604 and 0.615 when trained on the datasets of the years 2013 and 2014, respectively. Similarly, MamaDroid achieves average F1 scores of 0.579 and 0.614. These results can be attributed to whether the features selected by these methods can effectively capture the evolving characteristics of both malware and benign apps. For instance, despite the evolution of malware, the selected classification features of DroidSpan remained sustainable. 

In comparison, our approach performs better than the baseline in most years, and it achieves better average F1 scores overall, reaching 0.719 and 0.729. This result can be attributed to the analysis of behavior units, which allows for good robustness in the analysis of the program behaviors. In addition, our model has high generality because it only analyzes behavior sequences without the need to understand the high-level semantic information of behaviors related to Android systems. But our approach does not perform better than DroidSpan on the data of several years (e.g., 2014 and 2017) in Figure \ref{fig_7} (a). This shows that the analysis of DroidSpan for Android APPs is also robust. Since our approach lacks an understanding of the behavior features of Android (e.g., accessing sensitive data in Android), it does not perform as well as the DroidSpan on the data of several years. To improve the robustness of our approach to analyze the software behaviors of some specific operating systems (e.g., Android), the next step could be to explore combining behavior unit analysis with Android program behavior features, which we leave for future work.

\subsubsection{Generality Analysis}

To further investigate the effectiveness of our approach on other datasets, we extend our experiments to the ADFA-LD dataset \cite{creech2013generation}, which is also a widely used dataset for anomaly detection and consists of system call sequences. 
We use the same approach to generate adversarial samples in this dataset, as shown in Algorithm 2. For comparison, we benchmark our approach against other defense techniques. The experimental results are shown in Figure \ref{fig_8}. Our approach has the highest F1 scores on the data with different injection rates. When the attack injection rate increases from 5\% to 40\%, the F1 scores of our approach decrease by 5.68\% and 10.38\% on the two datasets, respectively. The decreases are the lowest among all the other approaches, which demonstrate the robustness of our approach across different datasets.
 
\section{Conclusion} \label{s:conclusion}
The RNN-based behavior analysis models are vulnerable to adversarial sample attacks. To mitigate this problem, this paper proposes an adversarial robust behavior sequence anomaly detection approach based on critical behavior unit learning. The perturbations of the irrelevant sequence are eliminated by identifying and extracting critical behavior units, and the robustness of the model is improved. A multi-level transformer-based abnormal behavior detection approach is proposed to learn the joint features within and between behavior units. The experimental results show that our proposed approach has good performance against obfuscation attacks.


%



\ifCLASSOPTIONcompsoc
  \section*{Acknowledgments}
\else
  \section*{Acknowledgment}
\fi

This work was supported by the National Key R\&D Program of China (No. 2021YFB2012402), the National Natural Science Foundation of China under Grants No. 61872111, and the Natural Science Foundation of Heilongjiang Province of China under Grants No. LH2023F017.

\ifCLASSOPTIONcaptionsoff
  \newpage
\fi



%



\bibliographystyle{IEEEtran}
\bibliography{sec}

\begin{thebibliography}{10}
\providecommand{\url}[1]{#1}
\csname url@samestyle\endcsname
\providecommand{\newblock}{\relax}
\providecommand{\bibinfo}[2]{#2}
\providecommand{\BIBentrySTDinterwordspacing}{\spaceskip=0pt\relax}
\providecommand{\BIBentryALTinterwordstretchfactor}{4}
\providecommand{\BIBentryALTinterwordspacing}{\spaceskip=\fontdimen2\font plus
\BIBentryALTinterwordstretchfactor\fontdimen3\font minus \fontdimen4\font\relax}
\providecommand{\BIBforeignlanguage}[2]{{%
\expandafter\ifx\csname l@#1\endcsname\relax
\typeout{** WARNING: IEEEtran.bst: No hyphenation pattern has been}%
\typeout{** loaded for the language `#1'. Using the pattern for}%
\typeout{** the default language instead.}%
\else
\language=\csname l@#1\endcsname
\fi
#2}}
\providecommand{\BIBdecl}{\relax}
\BIBdecl

\bibitem{afianian2019malware}
A.~Afianian, S.~Niksefat, B.~Sadeghiyan, and D.~Baptiste, ``Malware dynamic analysis evasion techniques: A survey,'' \emph{ACM Computing Surveys (CSUR)}, vol.~52, no.~6, pp. 1--28, 2019.

\bibitem{sahin2020survey}
M.~Sahin and S.~Bahtiyar, ``A survey on malware detection with deep learning,'' in \emph{13th International Conference on Security of Information and Networks}, 2020, pp. 1--6.

\bibitem{amin2020static}
M.~Amin, T.~A. Tanveer, M.~Tehseen, M.~Khan, F.~A. Khan, and S.~Anwar, ``Static malware detection and attribution in android byte-code through an end-to-end deep system,'' \emph{Future generation computer systems}, vol. 102, pp. 112--126, 2020.

\bibitem{li2022novel}
C.~Li, Q.~Lv, N.~Li, Y.~Wang, D.~Sun, and Y.~Qiao, ``A novel deep framework for dynamic malware detection based on api sequence intrinsic features,'' \emph{Computers \& Security}, vol. 116, p. 102686, 2022.

\bibitem{or2019dynamic}
O.~Or-Meir, N.~Nissim, Y.~Elovici, and L.~Rokach, ``Dynamic malware analysis in the modern era—a state of the art survey,'' \emph{ACM Computing Surveys (CSUR)}, vol.~52, no.~5, pp. 1--48, 2019.

\bibitem{forrest1996sense}
S.~Forrest, S.~A. Hofmeyr, A.~Somayaji, and T.~A. Longstaff, ``A sense of self for unix processes,'' in \emph{Proceedings 1996 IEEE Symposium on Security and Privacy}.\hskip 1em plus 0.5em minus 0.4em\relax IEEE, 1996, pp. 120--128.

\bibitem{firdausi2010analysis}
I.~Firdausi, A.~Erwin, A.~S. Nugroho \emph{et~al.}, ``Analysis of machine learning techniques used in behavior-based malware detection,'' in \emph{2010 second international conference on advances in computing, control, and telecommunication technologies}.\hskip 1em plus 0.5em minus 0.4em\relax IEEE, 2010, pp. 201--203.

\bibitem{vinayakumar2018detecting}
R.~Vinayakumar, K.~Soman, P.~Poornachandran, and S.~Sachin~Kumar, ``Detecting android malware using long short-term memory (lstm),'' \emph{Journal of Intelligent \& Fuzzy Systems}, vol.~34, no.~3, pp. 1277--1288, 2018.

\bibitem{liang2022adversarial}
H.~Liang, E.~He, Y.~Zhao, Z.~Jia, and H.~Li, ``Adversarial attack and defense: A survey,'' \emph{Electronics}, vol.~11, no.~8, p. 1283, 2022.

\bibitem{fang2022novel}
X.~Fang, Z.~Li, and G.~Yang, ``A novel approach to generating high-resolution adversarial examples,'' \emph{Applied Intelligence}, vol.~52, no.~2, pp. 1289--1305, 2022.

\bibitem{hu2018black}
W.~Hu and Y.~Tan, ``Black-box attacks against rnn based malware detection algorithms,'' in \emph{Workshops at the Thirty-Second AAAI Conference on Artificial Intelligence}, 2018.

\bibitem{fadadu2019evading}
F.~Fadadu, A.~Handa, N.~Kumar, and S.~K. Shukla, ``Evading api call sequence based malware classifiers,'' in \emph{International Conference on Information and Communications Security}.\hskip 1em plus 0.5em minus 0.4em\relax Springer, 2019, pp. 18--33.

\bibitem{rosenberg2020query}
I.~Rosenberg, A.~Shabtai, Y.~Elovici, and L.~Rokach, ``Query-efficient black-box attack against sequence-based malware classifiers,'' in \emph{Annual Computer Security Applications Conference}, 2020, pp. 611--626.

\bibitem{cai2020assessing}
H.~Cai, ``Assessing and improving malware detection sustainability through app evolution studies,'' \emph{ACM Transactions on Software Engineering and Methodology (TOSEM)}, vol.~29, no.~2, pp. 1--28, 2020.

\bibitem{cai2018droidcat}
H.~Cai, N.~Meng, B.~Ryder, and D.~Yao, ``Droidcat: Effective android malware detection and categorization via app-level profiling,'' \emph{IEEE Transactions on Information Forensics and Security}, vol.~14, no.~6, pp. 1455--1470, 2018.

\bibitem{szegedy2013intriguing}
C.~Szegedy, W.~Zaremba, I.~Sutskever, J.~Bruna, D.~Erhan, I.~Goodfellow, and R.~Fergus, ``Intriguing properties of neural networks,'' \emph{arXiv preprint arXiv:1312.6199}, 2013.

\bibitem{rosenberg2019defense}
I.~Rosenberg, A.~Shabtai, Y.~Elovici, and L.~\vspace{0mm}Rokach, ``Defense methods against adversarial examples for recurrent neural networks,'' \emph{arXiv preprint arXiv:1901.09963}, 2019.

\bibitem{rosenberg2021sequence}
I.~\vspace{0mm}Rosenberg, A.~Shabtai, Y.~Elovici, and L.~Rokach, ``Sequence squeezing: A defense method against adversarial examples for api call-based rnn variants,'' in \emph{2021 International Joint Conference on Neural Networks (IJCNN)}.\hskip 1em plus 0.5em minus 0.4em\relax IEEE, 2021, pp. 1--10.

\bibitem{hu2021using}
Z.~Hu, L.~Liu, H.~Yu, and X.~Yu, ``Using graph representation in host-based intrusion detection,'' \emph{Security and Communication Networks}, vol. 2021, 2021.

\bibitem{hasan2021megdroid}
H.~Hasan, B.~T. Ladani, and B.~Zamani, ``Megdroid: A model-driven event generation framework for dynamic android malware analysis,'' \emph{Information and Software Technology}, vol. 135, p. 106569, 2021.

\bibitem{d2021association}
G.~D’Angelo, M.~Ficco, and F.~Palmieri, ``Association rule-based malware classification using common subsequences of api calls,'' \emph{Applied Soft Computing}, vol. 105, p. 107234, 2021.

\bibitem{hardy2016dl4md}
W.~Hardy, L.~Chen, S.~Hou, Y.~Ye, and X.~Li, ``Dl4md: A deep learning framework for intelligent malware detection,'' in \emph{Proceedings of the International Conference on Data Science (ICDATA)}.\hskip 1em plus 0.5em minus 0.4em\relax The Steering Committee of The World Congress in Computer Science, Computer~…, 2016, p.~61.

\bibitem{rhode2018early}
M.~Rhode, P.~Burnap, and K.~Jones, ``Early-stage malware prediction using recurrent neural networks,'' \emph{computers \& security}, vol.~77, pp. 578--594, 2018.

\bibitem{natani2013malware}
P.~Natani and D.~Vidyarthi, ``Malware detection using api function frequency with ensemble based classifier,'' in \emph{International Symposium on Security in Computing and Communication}.\hskip 1em plus 0.5em minus 0.4em\relax Springer, 2013, pp. 378--388.

\bibitem{xiao2019android}
X.~Xiao, S.~Zhang, F.~Mercaldo, G.~Hu, and A.~K. Sangaiah, ``Android malware detection based on system call sequences and lstm,'' \emph{Multimedia Tools and Applications}, vol.~78, no.~4, pp. 3979--3999, 2019.

\bibitem{guan2021malware}
Y.~Guan and N.~Ezzati-Jivan, ``Malware system calls detection using hybrid system,'' in \emph{2021 IEEE International Systems Conference (SysCon)}.\hskip 1em plus 0.5em minus 0.4em\relax IEEE, 2021, pp. 1--8.

\bibitem{rosenberg2018generic}
I.~Rosenberg, A.~Shabtai, L.~Rokach, and Y.~Elovici, ``Generic black-box end-to-end attack against state of the art api call based malware classifiers,'' in \emph{International Symposium on Research in Attacks, Intrusions, and Defenses}.\hskip 1em plus 0.5em minus 0.4em\relax Springer, 2018, pp. 490--510.

\bibitem{madry2017towards}
A.~Madry, A.~Makelov, L.~Schmidt, D.~Tsipras, and A.~Vladu, ``Towards deep learning models resistant to adversarial attacks,'' \emph{arXiv preprint arXiv:1706.06083}, 2017.

\bibitem{apruzzese2022real}
G.~Apruzzese, H.~S. Anderson, S.~Dambra, D.~Freeman, F.~Pierazzi, and K.~A. Roundy, ``" real attackers don't compute gradients": Bridging the gap between adversarial ml research and practice,'' \emph{arXiv preprint arXiv:2212.14315}, 2022.

\bibitem{ye2009time}
L.~Ye and E.~Keogh, ``Time series shapelets: a new primitive for data mining,'' in \emph{Proceedings of the 15th ACM SIGKDD international conference on Knowledge discovery and data mining}, 2009, pp. 947--956.

\bibitem{lines2012shapelet}
J.~Lines, L.~M. Davis, J.~Hills, and A.~Bagnall, ``A shapelet transform for time series classification,'' in \emph{Proceedings of the 18th ACM SIGKDD international conference on Knowledge discovery and data mining}, 2012, pp. 289--297.

\bibitem{grabocka2014learning}
J.~Grabocka, N.~Schilling, M.~Wistuba, and L.~Schmidt-Thieme, ``Learning time-series shapelets,'' in \emph{Proceedings of the 20th ACM SIGKDD international conference on Knowledge discovery and data mining}, 2014, pp. 392--401.

\bibitem{bergroth2000survey}
L.~Bergroth, H.~Hakonen, and T.~Raita, ``A survey of longest common subsequence algorithms,'' in \emph{Proceedings Seventh International Symposium on String Processing and Information Retrieval. SPIRE 2000}.\hskip 1em plus 0.5em minus 0.4em\relax IEEE, 2000, pp. 39--48.

\bibitem{al2019use}
S.~Al-Saqqa and A.~Awajan, ``The use of word2vec model in sentiment analysis: A survey,'' in \emph{Proceedings of the 2019 International Conference on Artificial Intelligence, Robotics and Control}, 2019, pp. 39--43.

\bibitem{du2017deeplog}
M.~Du, F.~Li, G.~Zheng, and V.~Srikumar, ``Deeplog: Anomaly detection and diagnosis from system logs through deep learning,'' in \emph{Proceedings of the 2017 ACM SIGSAC conference on computer and communications security}, 2017, pp. 1285--1298.

\bibitem{nedelkoski2020self}
S.~Nedelkoski, J.~Bogatinovski, A.~Acker, J.~Cardoso, and O.~Kao, ``Self-attentive classification-based anomaly detection in unstructured logs,'' in \emph{2020 IEEE International Conference on Data Mining (ICDM)}.\hskip 1em plus 0.5em minus 0.4em\relax IEEE, 2020, pp. 1196--1201.

\bibitem{ludetecting}
S.~Lu, X.~Wei, Y.~Li, and L.~Wang, ``Detecting anomaly in big data system logs using convolutional neural network. in 2018 ieee 16th intl conf on dependable, autonomic and secure computing, 16th intl conf on pervasive intelligence and computing,'' in \emph{4th Intl Conf on Big Data Intelligence and Computing and Cyber Science and Technology Congress (DASC/PiCom/DataCom/CyberSciTech)}, pp. 151--158.

\bibitem{farzad2020unsupervised}
A.~Farzad and T.~A. Gulliver, ``Unsupervised log message anomaly detection,'' \emph{ICT Express}, vol.~6, no.~3, pp. 229--237, 2020.

\bibitem{li2021androct}
W.~Li, X.~Fu, and H.~Cai, ``Androct: ten years of app call traces in android,'' in \emph{2021 IEEE/ACM 18th International Conference on Mining Software Repositories (MSR)}.\hskip 1em plus 0.5em minus 0.4em\relax IEEE, 2021, pp. 570--574.

\bibitem{yu2017seqgan}
L.~Yu, W.~Zhang, J.~Wang, and Y.~Yu, ``Seqgan: Sequence generative adversarial nets with policy gradient,'' in \emph{Proceedings of the AAAI conference on artificial intelligence}, vol.~31, no.~1, 2017.

\bibitem{das2015semantics}
S.~Das, Y.~Liu, W.~Zhang, and M.~Chandramohan, ``Semantics-based online malware detection: Towards efficient real-time protection against malware,'' \emph{IEEE transactions on information forensics and security}, vol.~11, no.~2, pp. 289--302, 2015.

\bibitem{cai2020embracing}
H.~Cai, ``Embracing mobile app evolution via continuous ecosystem mining and characterization,'' in \emph{Proceedings of the IEEE/ACM 7th International Conference on Mobile Software Engineering and Systems}, 2020, pp. 31--35.

\bibitem{xu2019droidevolver}
K.~Xu, Y.~Li, R.~Deng, K.~Chen, and J.~Xu, ``Droidevolver: Self-evolving android malware detection system,'' in \emph{2019 IEEE European Symposium on Security and Privacy (EuroS\&P)}.\hskip 1em plus 0.5em minus 0.4em\relax IEEE, 2019, pp. 47--62.

\bibitem{onwuzurike2019mamadroid}
L.~Onwuzurike, E.~Mariconti, P.~Andriotis, E.~D. Cristofaro, G.~Ross, and G.~Stringhini, ``Mamadroid: Detecting android malware by building markov chains of behavioral models (extended version),'' \emph{ACM Transactions on Privacy and Security (TOPS)}, vol.~22, no.~2, pp. 1--34, 2019.

\bibitem{creech2013generation}
G.~Creech and J.~Hu, ``Generation of a new ids test dataset: Time to retire the kdd collection,'' in \emph{2013 IEEE Wireless Communications and Networking Conference (WCNC)}.\hskip 1em plus 0.5em minus 0.4em\relax IEEE, 2013, pp. 4487--4492.

\end{thebibliography}

%

\begin{IEEEbiography}[{\includegraphics[width=1in,height=1.25in,clip,keepaspectratio]{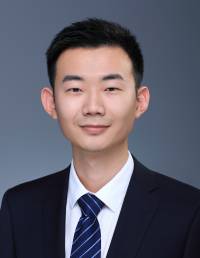}}]{Dongyang Zhan} is an assistant professor in School of Cyberspace Science at Harbin Institute of Technology. He received the B.S. degree in Computer Science from Harbin Institute of Technology from 2010 to 2014. From 2015 to 2019, he has been working as a Ph.D. candidate in School of Computer Science and Technology at HIT. His research interests include cloud computing and security.
\end{IEEEbiography}

\begin{IEEEbiography}[{\includegraphics[width=1in,height=1.25in,clip,keepaspectratio]{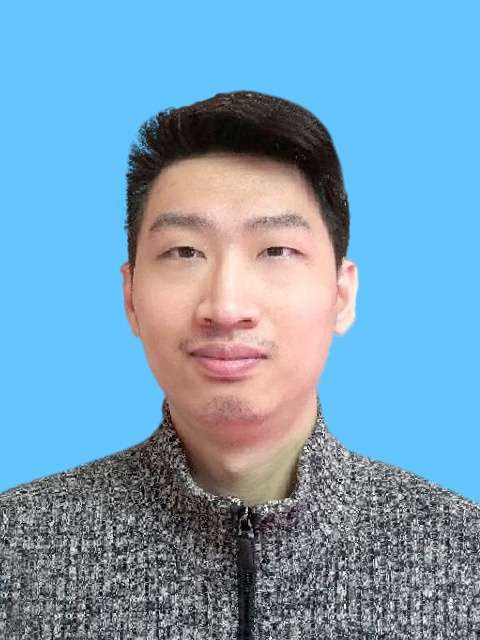}}]{Kai Tan} is a PhD candidate from the Harbin Institute of Technology, China. His research focuses on cloud security.
\end{IEEEbiography}

\begin{IEEEbiography}[{\includegraphics[width=1in,height=1.25in,clip,keepaspectratio]{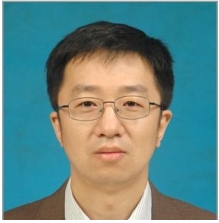}}]{Xiangzhan Yu}
is a professor in School of Cyberspace Science at Harbin Institute of Technology. His main research fields include: network and information security, security of internet of things and privacy protection. He has published one academic book and more than 50 papers on international journals and conferences.
\end{IEEEbiography}

\begin{IEEEbiography}[{\includegraphics[width=1in,height=1.25in,clip,keepaspectratio]{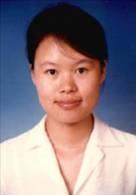}}]{Hongli Zhang}
received her BS degree in Computer Science from Sichuan University, Chengdu, China in 1994, and her Ph.D. degree in Computer Science from Harbin Institute of Technology (HIT), Harbin, China in 1999. She is currently a Professor in School of Cyberspace Science in HIT. Her research interests include network and information security, network measurement and modeling, and parallel processing.
\end{IEEEbiography}


\begin{IEEEbiography}[{\includegraphics[width=1in,height=1.25in,clip,keepaspectratio]{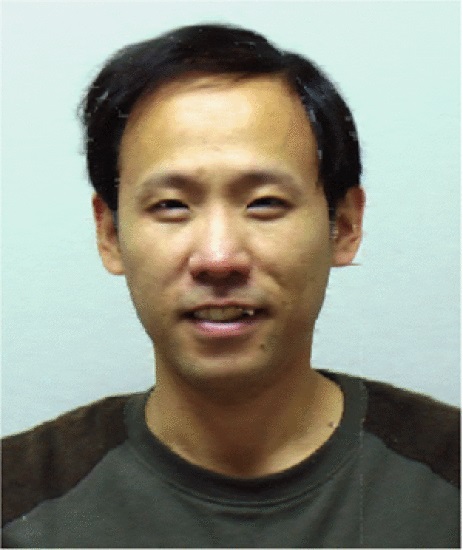}}]{Lin Ye}
received the Ph.D. degree at Harbin Institute of Technology in 2011. From January 2016 to January 2017, he was a visiting scholar in the Department of Computer and Information Sciences, Temple University, USA. His current research interests include network security, peer-to-peer network, network measurement and cloud computing.
\end{IEEEbiography}

\begin{IEEEbiography}[{\includegraphics[width=1in,height=1.25in,clip,keepaspectratio]{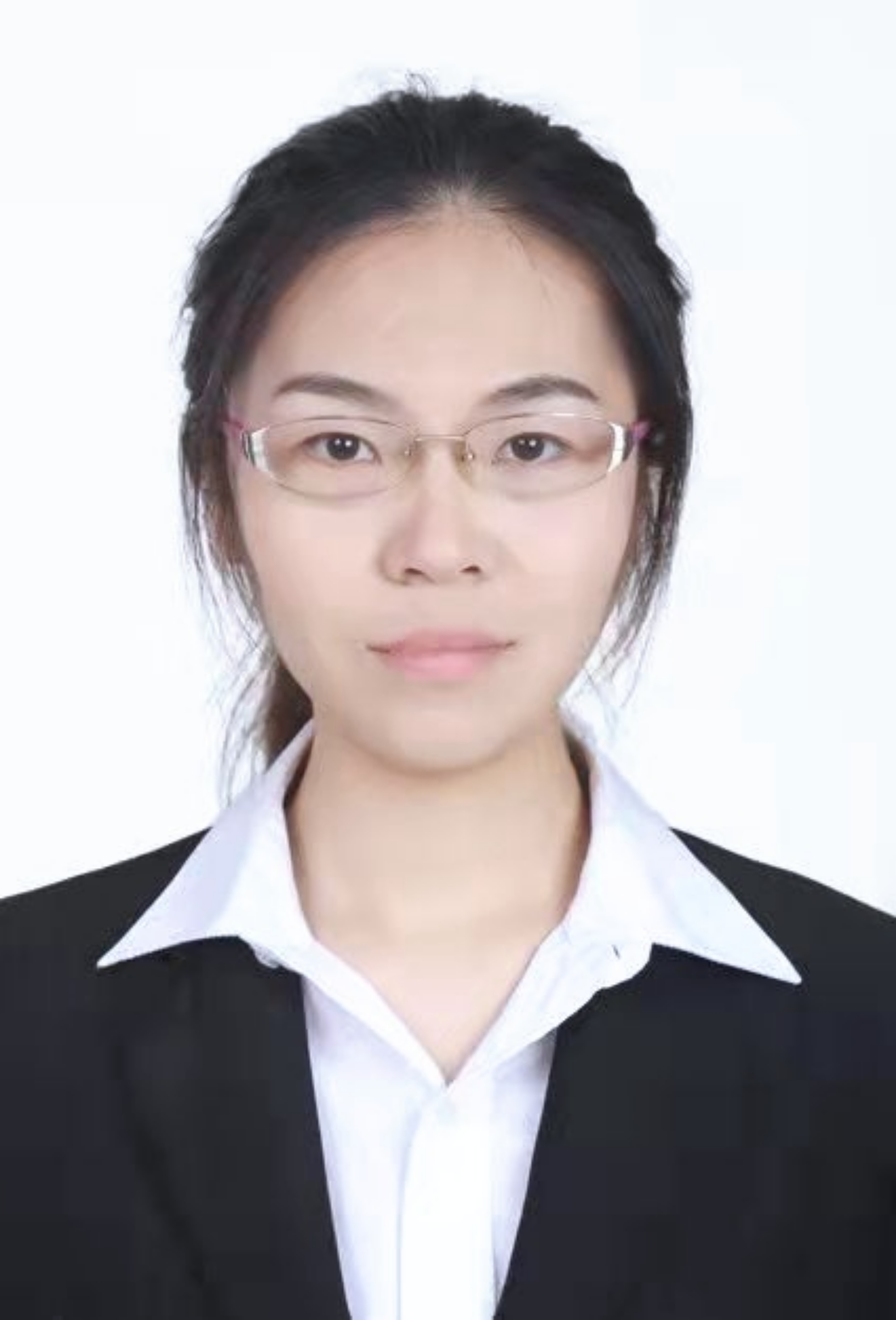}}]{Zheng He} is an engineer in Heilongjiang Meteorological Bureau. She received her bachelor's and Master’s degrees in Meteorology Science in Nanjing University of Information Science and Technology from 2011 to 2018. From 2018, she has been working in Weather Modification Office of Heilongjiang Province. Her research interests include climate change, weather modification and machine learning.
\end{IEEEbiography}




\end{document}